\documentclass[preprint2,longabstract]{aastex}

\newcommand\lam{$\lambda$}
\newcommand{\kms}{km~s$^{-1}$}

\newcommand\OI{[O\,{\sc i}]}
\newcommand\OII{[O\,{\sc ii}]}
\newcommand\OIII{[O\,{\sc iii}]}
\newcommand\NII{[N\,{\sc ii}]}
\newcommand\SII{[S\,{\sc ii}]}
\newcommand\Ha{H\,$\alpha$}
\newcommand\Hb{H\,$\beta$}
\newcommand\Hg{H\,$\gamma$}
\newcommand\degree{$^\circ$}
\newcommand\ergss{ergs s$^{-1}$}

\begin{document}

\title{The Dynamics of Abell 2125}

\author{Neal A. Miller} 
\affil{NASA Goddard Space Flight Center\\
Laboratory for Astronomy \& Solar Physics, Code 681\\
Greenbelt, MD \ 20771}
\email{nmiller@stis.gsfc.nasa.gov}

\author{Frazer N. Owen\altaffilmark{2}}
\affil{National Radio Astronomy Observatory\altaffilmark{1}\\
P.O. Box O, \\
Socorro, New Mexico 87801}
\email{fowen@aoc.nrao.edu}

\author{John M. Hill\altaffilmark{2}}
\affil{Large Binocular Telescope Observatory\\
University of Arizona \\
Tucson, AZ \ 85721}
\email{jhill@as.arizona.edu}

\author{William C. Keel\altaffilmark{2}}
\affil{Department of Physics and Astronomy\\
University of Alabama\\
Tuscaloosa, AL 35487}
\email{keel@bildad.astr.ua.edu}

\author{Michael J. Ledlow\altaffilmark{2}}
\affil{Gemini Observatory\\
c/o AURA, Casilla 603\\
La Serena, Chile}
\email{mledlow@gemini.edu}

\author{William R. Oegerle\altaffilmark{2}}
\affil{NASA Goddard Space Flight Center\\
Laboratory for Astronomy \& Solar Physics, Code 680\\
Greenbelt, MD \ 20771}
\email{william.r.oegerle@nasa.gov}

\altaffiltext{1}{The National Radio Astronomy Observatory is a facility of the National Science Foundation operated under cooperative agreement by Associated Universities, Inc.}

\altaffiltext{2}{Visiting Astronomer, Kitt Peak National Observatory, National Optical Astronomy Observatories, which is operated by the Association of Universities for Research in Astronomy, Inc., under cooperative agreement with the National Science Foundation.}

\vspace{1.0in}

\begin{abstract} 

We present 371 galaxy velocities in the field of the very rich cluster Abell 2125 ($z\approx0.25$). These were determined using optical spectroscopy collected over several years from both the WIYN 3.5m telescope and NOAO Mayall 4m telescope. Prior studies at a variety of wavelengths (radio, optical, and X-ray) have indicated that A2125 is a likely cluster-cluster merger, a scenario which we are able to test using our large velocity database. We identified 224 cluster galaxies, which were subjected to a broad range of statistical tests using both positional and velocity information to evaluate the cluster dynamics and substructure. The tests confirmed the presence of substructures within the Abell 2125 system at high significance, demonstrating that A2125 is a complex dynamical system. Comparison of the test results with existing simulations strengthens the merger hypothesis, and provides clues about the merger geometry and stage. The merger model for the system can reconcile A2125's low X-ray temperature and luminosity with its apparently high richness, and might also explain A2125's high fraction of active galaxies identified in prior radio and optical studies.

\end{abstract}

\section{Introduction}

Abell 2125 is a very rich cluster of galaxies \citep[Richness Class 4;][]{aco1989} found at intermediate redshift ($z\approx0.25$). It was identified as an active cluster in the \citet{butc1984} study, with a noted blue fraction  of $0.19$. This would suggest an increase in the star formation of A2125 galaxies relative to those found in low redshift clusters, which typically have blue fractions $\sim0.03$. Considerable effort has been expended to confirm this suggestion via radio continuum observations \citep{dwar1999,owen2003}. The radio continuum emission at 20 cm for normal and starbursting spirals is strongly correlated with far-infrared (FIR) emission \citep[e.g.,][]{helo1985,yun2001}. It is believed that the source of this correlation is massive stars, whose supernovae accelerate electrons which radiate synchrotron emission in galactic magnetic fields \citep[for a review, see][]{cond1992}. The same massive stars are the dominant source of FIR emission as they heat gas and dust in their parent molecular clouds. Indeed, analogous to the photometric results, A2125 was shown to have a large radio galaxy fraction implying an elevated level of star formation \citep{dwar1999}. A companion study adding optical spectroscopy confirmed that the identified radio galaxies were cluster members, and argued that the high blue/radio fraction was due to the dynamical state of A2125, which appears to be that of a cluster-cluster merger \citep{owen1999}. Additional radio and
optical observations have extended this result to much lower luminosities \citep{owen2003}, further underscoring the high level of activity in A2125. This activity includes a high fraction of active galactic nuclei, as well as an excess of star-forming galaxies.

The interpretation of A2125 as a cluster-cluster merger is based on both its galaxy distribution \citep{owen1999} and its X-ray structure \citep{wang1997,wang2004}. The \citet{wang1997} X-ray observations were performed with the {\it ROSAT} PSPC, and indicated a complex environment. This impression was strengthened by recent {\it Chandra} observations in \citet{wang2004}. Multiple peaks in the X-ray emission were identified, in addition to larger scale diffuse emission. The emission coinciding with the core of A2125 was shown to have an ellipticity of 0.25--0.33 ({\it Chandra} results, $90\%$ confidence range) and a north-to-east position angle between 115\degree{} and 127\degree. The X-ray emission corresponding to the core of A2125 was also noted to be clumpy. In conjunction with the low measured temperature of 3.2 keV, these findings indicated that A2125 was a dynamically complex system. Additional peaks in the X-ray emission outside the main body of A2125 were observed to the north, along with more diffuse extended emission to the southwest. This diffuse emission is of cooler temperature ($\sim$1 keV), and correlates with concentrations of spiral galaxies. It might thereby represent diffuse intergalactic gas associated with large-scale hierarchical structure formation. In total, the X-ray observations led to the interpretation of the entire system as a supercluster over 7 Mpc in extent. Interestingly, the X-ray ellipticity of the main body of A2125 was roughly perpendicular to the overall superstructure, leading the authors to suggest that the core of A2125 represented ``an initial coalescence of subunits of comparable mass.'' 

Consequently, there is much interest in a detailed dynamical assessment of A2125. The prior studies lacked large numbers of measured galaxy velocities with  which to assess possible substructures and the overall merger environment, and we have undertaken a broad campaign of optical spectroscopy to remedy this. In this paper, we report over 450 new velocity measurements of 371 galaxies in the field of A2125. The optical spectra were collected using several instruments, including the Hydra multifiber spectrograph on the WIYN 3.5m telescope, multislit spectra collected using the Mayall 4m telescope, and long-slit spectra taken with the RC spectrograph at the Mayall telescope \citep[previously published in][]{owen1999}. The galaxy velocities and positions were subjected to a collection of statistical tests in order to identify substructure in A2125 and quantify its significance. A2125 is shown to be a highly complex system, rich in substructure. Comparison of the test results with those determined for simulations indicates that the data are consistent with models of cluster-cluster mergers. In this interpretation, A2125 is viewed along an axis offset slightly from the merger axis and at an epoch within a few hundred million years of the time when the cores of the merging clusters have crossed.

We will begin with a description of the spectroscopy database in Section \ref{sec-data}. Analysis of the resulting velocities is presented in Section \ref{sec-analysis}. In Section \ref{sec-discuss} we assess the merger hypothesis by comparing the substructure tests with computer models, and view the results in light of X-ray and radio investigations of the cluster. Our conclusions are briefly summarized in Section \ref{sec-conclude}. We have adopted the WMAP cosmology of $H_0 = 71$ \kms~Mpc$^{-1}$, $\Omega_m=0.27$, and $\Omega_\Lambda=0.73$ for all distance-dependent calculations, meaning 1\arcsec{} corresponds to about 3.8 kpc at A2125. The classical Abell radius \citep[$R_A \equiv 1.7^\prime / z$;][]{abell1958} for the cluster is thereby 1.59 Mpc.

\section{Observations and Data Reductions}\label{sec-data}

\subsection{Multifiber Spectroscopy using WIYN with Hydra}\label{sec-hydra}

\subsubsection{1996 Observations}

Observations of A2125 were performed by WRO and JMH using the Hydra multifiber spectrograph on the WIYN 3.5m telescope during a pair of observing runs in 1996. Three fiber configurations were observed during 16--18 April 1996, and a fourth was observed during 20--22 May 1996. Each run used the blue fibers (3.1\arcsec) and a 400 lines mm$^{-1}$ grating blazed at 4000\mbox{\AA}. This produced wavelength coverage from about 4300\mbox{\AA} to 7500\mbox{\AA} at a resolution of 7\mbox{\AA}. Each configuration was observed multiple times to reject
cosmic rays and produce high signal-to-noise spectra. Total exposure times varied depending on observing conditions, but ranged from three to five hours (see the summary in Table \ref{tbl-obs}).

Calibration frames were taken using standard methods. At least 20 bias frames were collected on each night, as well as at least three flat fields per fiber configuration. Wavelength calibration was achieved via observation of CuAr lamps, performed both before and after science observations of a given fiber configuration. The wavelength solutions typically produced an rms residual less than 0.15\mbox{\AA} as determined from about 45 lines. Of the 96 available fibers per configuration, typically around 30 were assigned to random locations to produce a good quality sky spectrum.

Each science exposure was reduced individually, with the set of exposures for a given fiber configuration then combined to produce the final spectra. The standard IRAF tasks within the NOAO {\scshape hydra} package were used to produce the final wavelength-calibrated, sky-subtracted spectra for each exposure. The set of exposures corresponding to a given fiber configuration were then median combined after weighting by their modes (as defined in a region between 5000\mbox{\AA} and 5500\mbox{\AA}, chosen to avoid strong emission and absorption features in cluster galaxies). 

Measurement of galaxy velocities proceeded along one of two paths. First, the spectra were cross correlated against a collection of velocity standards \citep[using the IRAF {\scshape rv} package; see][]{tonr1979}. These included spectra of the nearby galaxies NGC 3379 (assumed velocity 922 \kms), NGC 7331 (819 \kms), M31 (-297 \kms), M32 (-200 \kms), and a bright elliptical in A779 (6877 \kms, determined from cross correlation with the other templates and observed multiple times), plus standard stars HD 52071 (94.7 \kms) and BD +25 3190 (-49.5 \kms). The velocity standard spectra (hereafter ``template'' spectra) were collected over several years using a variety of spectrographs at resolutions comparable to the present observations. In total, the template list included 22 spectra. Cross correlation velocities for each observed galaxy were then determined from the velocities derived using each of the 22 templates. First, the average of the velocities was determined \citep[weighted by the $R$ values, where $R$ is a measure of the goodness of fit defined as the ratio of the height of the peak in the cross correlation to that of the average peak; see][]{tonr1979} and outliers were clipped ($\pm3\sigma$). The dispersion in the remaining velocities was required to be less than 115 \kms, and we also required that the $R$ values were greater than 2.5. Finally, we required that at least $60\%$ of the templates produced acceptable velocities given these restrictions. Initial errors were based on the weighted average $R$ value, such that the error equals $280(1+R)^{-1}$ \kms{} \citep[see][]{tonr1979,hill1998}. The final error represents this value added in quadrature with an additional uncertainty of 40 \kms{} (see Section \ref{sec-comb} below).

The spectra were also inspected by eye by two of us (NAM and FNO) independently to confirm the results. In addition to confirmation of cross correlation velocities, many emission line spectra were identified. For these galaxies, we fit Gaussian profiles to up to eleven emission lines typically observed in star-forming galaxies and AGN: \OII{} 3727\mbox{\AA}, \Hg{} 4340\mbox{\AA}, \Hb{} 4861\mbox{\AA}, \OIII{} 4959\mbox{\AA}, \OIII{} 5007\mbox{\AA}, \OI{} 6300\mbox{\AA}, \NII{} 6548\mbox{\AA}, \Ha{} 6563\mbox{\AA}, \NII{} 6583\mbox{\AA}, \SII{} 6716\mbox{\AA}, and \SII{} 6731\mbox{\AA}. Velocities were determined for each detected line, and a net galaxy velocity was derived from these via a simple average (requiring at least two emission lines be measured). The heliocentric correction was then applied to arrive at the final emission line velocity. To evaluate the errors, we determined the dispersion in the line measurements for each galaxy. As this quantity is somewhat dependent on the number of measured lines (e.g., a galaxy with only two measured lines may have a much lower associated dispersion than a strong emission line galaxy with ten lines), we calculated the average dispersion over all emission line galaxies, arriving at a value of 39 \kms. This represents the combination of various error terms including: errors in measurement of line positions, wavelength calibration errors (which should be about 10 \kms{} based on the rms residual obtained from the wavelength solution), and gas motions within the galaxies. We therefore adopted 40 \kms{} as the error in our velocities determined from emission lines.

Final velocities were then adopted based on the cross correlation and emission line values. For galaxies proving to be cluster members (see below), these observed heliocentric corrected velocities ($cz$) may be found in Table \ref{tbl-vels}. For non-cluster members (foreground and background galaxies and quasars), the heliocentric redshifts may be found in Table \ref{tbl-xvels}. In general, we used the emission line values whenever they were available. For several cases in which the lines were weak or noisy, we instead used the cross correlation values. These are marked in Tables \ref{tbl-vels} and \ref{tbl-xvels} with the noted emission lines offset in parentheses. Note that in nearly all cases where both an emission line and a cross correlation velocity were available, the two velocities were consistent within about 150 \kms. A small number of cross correlation velocities with low $R$ value were adopted in specific cases where a single emission line measurement (usually \OII) confirmed the cross correlation value.

\subsubsection{2000 Observations}

Additional spectroscopy using Hydra on the WIYN telescope was performed in the queue observing mode over several nights from March through May 2000 (see Table \ref{tbl-obs}). These used the red cables (2\arcsec{} fibers) and the 316 lines mm$^{-1}$ grating blazed at 7\degree, providing wavelength coverage from $\sim4300\mbox{\AA}$ to $\sim9150\mbox{\AA}$ at $\sim6\mbox{\AA}$ resolution. Three fiber configurations each consisting of $\sim65$ target galaxies were observed in five or six exposures each of 2800 seconds duration. Most of the remaining fibers were used to create a sky spectrum for data reduction. In addition to the science exposures, CuAr wavelength calibration frames were interspersed along with observations of velocity and spectrophotometric standards. Zeros, dome flats, and sky flats were also obtained on each night.

The data reduction and determination of velocities was similar to that for the 1996 data. The procedure for determining the cross correlation velocities differed slightly from that outlined above. Five template spectra, corresponding to M31, M32, NGC 3779, NGC 7331, and HD132737 (assumed velocity of -24 \kms) were each run interactively with the individual spectra. Regions for the correlations were chosen on a case-by-case basis to emphasize stronger absorption features such as the 4000$\mbox{\AA}$ break, with the best fit (i.e., highest $R$ value) being adopted as the velocity for a given galaxy. Velocity errors were then determined in the same manner as for the 1996 WIYN data.

\subsection{Multislit Spectroscopy at 4m Mayall Telescope}

The 4m Mayall Telescope was used for spectroscopic observations of A2125 by WCK, FNO, and MJL on three nights from 27--29 May 1998 (Table \ref{tbl-obs}). These observations were performed using 13 custom-made multislit masks, each consisting of between eight and thirteen 2.5\arcsec{} slits centered on target galaxies. The primary motivation of these observations was to obtain spectra of galaxies with radio emission \citep[see][]{owen2003}, with additional slits placed on galaxies with $m_R \leq 20$. The RC spectrograph was used along with grating KPC-10A (316 lines mm$^{-1}$, blazed at 4000$\mbox{\AA}$), providing a resolution of $\sim7\mbox{\AA}$ and coverage from 4000$\mbox{\AA}$ past 9200$\mbox{\AA}$. Each mask was observed for a total exposure time of one hour, split into two 30-minute exposures to facilitate rejection of cosmic rays. Flux calibration was achieved through observation of standard stars through slits of the same width, including standard stars viewed through slits from the actual science multislit masks. Data reduction and determination of galaxy velocities were achieved in the same manner as that applied to the 2000 WIYN data.

\subsection{Additional Data and Combination of Velocities}\label{sec-comb}

In addition to the velocity data described above, we have included the velocity data from \citet{owen1999}. These were obtained using a long slit on the KPNO Mayall 4m telescope with the RC spectrograph. Most of these galaxies have been re-observed by at least one of the above spectroscopic campaigns, meaning only three velocities (two of which are cluster galaxies) are added to the data from this source. In total, there are 374 galaxies in our database including 371 listed in Tables \ref{tbl-vels} and \ref{tbl-xvels}. Many galaxies have spectroscopy results from multiple observing runs, meaning the two tables contain a total of 454 entries.

Visual inspection of the spectra also identified a handful of quasars. These have been noted in Table \ref{tbl-xvels}, along with their redshifts and comments indicating the emission features upon which such measurements are based. The broad emission lines make precise determination of redshifts difficult, so we have arbitrarily assigned an error of $\Delta z=0.001$ (300 \kms) to these redshifts.

With many galaxies being the targets of multiple observing runs, we explored possible offsets in our derived velocities. We adopted the 1996 WIYN Hydra observations as a benchmark, and compared the velocities obtained from these observations with the velocities obtained from the other observations (the 2000 WIYN Hydra observations and the 1998 Mayall observations). These comparisons were further split into velocities determined by cross correlation and velocities determined by locations of emission lines. The emission line velocities in common were consistent for each data set (mean difference between 2000 WIYN data and 1996 WIYN data of -7 \kms{} for 17 galaxies, mean difference between 1998 Mayall data and 1996 WIYN data of 12 \kms{} for 10 galaxies). The velocities determined via cross correlation were also generally consistent, with the 2000 WIYN velocities an average of 51 \kms{} greater than the 1996 WIYN velocities (17 objects in common) and the 1998 Mayall velocities an average of 15 \kms{} less than the 1996 WIYN velocities (based on 24 common objects). Differences as small as these can easily result from sampling different portions of the target galaxies through combinations of astrometry, pointing, and aperture size.

We further tested possible offsets using velocity standard star observations. Typically, a velocity standard star was observed on each night of the observing runs. Velocities for these standard stars were determined via cross correlation using the various template velocity standard spectra. The mean velocity for a given standard star, determined from the full set of template spectra, was usually within 15 \kms{} of the cataloged velocity. The full spread in 
velocities determined from the various templates was $\lesssim70$ \kms, with a typical dispersion around 20 \kms. Rather than apply any zero point shifts to velocities determined from data collected on specific nights and observing runs (none of which would be statistically significant), we have elected to include an additional error of 40 \kms{} (added in quadrature) to the reported galaxy velocities.

In compiling our final velocity list for dynamical analysis, we prioritized the observations as follows. If a velocity obtained from the 1996 WIYN observations was available for a given galaxy, it was adopted. If a galaxy was unobserved by the 1996 WIYN observations, we used its 2000 WIYN velocity (if available), followed by any 1998 Mayall velocity and lastly by any \citet{owen1999} velocity. This hierarchy is largely governed by numbers and is meant to maintain as consistent a velocity database as possible.

With spectroscopy data coming from four separate sources, it is useful to investigate possible bias in our sampling of cluster galaxies (e.g., obtaining relatively more velocities for some portions of the cluster than for others). Fortunately, this is mitigated by the large (60\arcmin{} diameter) field of view of WIYN+Hydra which equates to nearly 7 Mpc at the redshift of the cluster. In Figure \ref{fig-allpos} we plot the positions of all galaxies for which we have spectra. It can be seen that our spatial sampling is relatively uniform, particularly by noting the distribution of background galaxies (open triangles in the figure). Optical images of the cluster, particularly those with high resolution, often reveal tight groupings of galaxies \citep{owen2003}. Limitations on fiber spacing with Hydra (37\arcsec) mean that any single fiber configuration will miss some galaxies. However, since seven different fiber configurations were observed (four in 1996, three in 2000) this effect is lessened. Overall, the net effect of missing a few galaxies in tight groupings should be minimal in performing the dynamical assessments. In fact, missing spectra for some galaxies in tight groups would result in less pronounced evidence for substructure than is actually present.

\section{Analysis: Tests of Substructure}\label{sec-analysis}

A histogram of the velocity data is presented in Figure \ref{fig-vhfull}. A2125 is easily seen as the single strong peak in the velocity distribution. In all, there are 224 cluster galaxies as identified via 3$\sigma$ clipping of the velocity data \citep{yahi1977}. The calculated biweight location and scale \citep[see][]{beer1990} are $73897\pm74$ \kms{} ($z=0.2465$) and $1113^{+57}_{-49}$ \kms, respectively. These numbers have been corrected for measurement errors \citep[see][]{dane1980}, and the dispersion was calculated from relativistically-corrected velocities (which we use for all subsequent dynamics calculations). We adopt the biweight location as our cluster systemic velocity and the biweight scale as our cluster velocity dispersion for the remainder of this discussion. Note that the ``non-robust'' calculations of the mean and dispersion yield $73995\pm96$ \kms{} and $1150^{+59}_{-51}$ \kms, similar to the robust quantities.

The 224 cluster galaxies were subjected to a battery of statistical tests to evaluate potential substructure. These included tests performed exclusively on the velocity data \citep[the ``ROSTAT'' package of][]{beer1990}, and tests on the galaxy positional data both with and without velocity information \citep[for a complete listing and description of all tests, see][hereafter PRBB]{pink1996}. The advantages to using a battery of statistical tests are noted in PRBB. Essentially, each test provides a different perspective on the dynamical state of the cluster and taken collectively they often complement one another. Understanding a set of results is aided greatly by comparison with numerous N-body simulations.

The tests performed exclusively on the velocity data (``1D'' tests) generally compare the distribution to that of a Gaussian, and are presented in Table \ref{tbl-1Dsig}. Since the computational algorithms in the ROSTAT package which are used to determine the significance for the $B_1$ and $B_2$ statistics ($B_1$ and $B_2$ are the canonical estimators of skewness and kurtosis, and represent the third and fourth moments of the velocity distribution) fail for large $N$, we have estimated their significance via 40 randomly-drawn subsets (with $N=150$) of the full data set. Most of the 1D tests indicated significant non-Gaussian behavior of the velocity histogram. Qualitatively, this may be understood by inspection of Figure \ref{fig-vhfull}. In particular, the cluster histogram has a ``tail'' at higher velocity. This produces significant skewness to the distribution, but does not make it overly kurtotic (e.g., the cluster distribution is more skewed than a Gaussian at better than $95\%$ confidence based on the $B_1$ statistic, yet consistent with a Gaussian in the sense of kurtosis based on the $B_2$ statistic). Consistent with the visual interpretation, combinations of the statistics indicate the distribution has more counts to the higher velocity side of the mean and similarly that the long tail identified by the tests resides on this side.

Similarly, tests performed on the positional data both with and without the associated velocities produced strong evidence for substructure. Table \ref{tbl-MDsig} presents the nine tests performed and their results. With the exceptions of the Angular Separation Test \citep{west1988} and the $\epsilon$ test \citep{bird1993}, all indicate highly significant substructure (at $\geq99.7\%$ confidence, as determined via 1000 Monte Carlo simulations). As an example, in Figure \ref{fig-dstest} we present the graphical representation of the $\Delta$ test \citep{dres1988}. This test calculates local deviations from the systemic velocity ($v_{sys}$) and dispersion ($\sigma$) of the entire cluster:
\begin{equation}\label{eqn-ds}
\delta_i ^2 = \frac{N}{\sigma ^2} \left[ (v_{local} - v_{sys})^2 + 
(\sigma_{local} - \sigma)^2 \right].
\end{equation}
Here $N$ is the number of galaxies which defines the local environment around galaxy $i$, taken to be $N=11$ in the original formulation of the test (i.e., a given galaxy and its ten nearest neighbors). For our calculations, we adopt $N=15$ (the square root of the total number of galaxies) as in PRBB. The sum of $\delta$ over all cluster members represents the actual test statistic, $\Delta$, and is approximately equal to the total number of cluster galaxies in the case of no substructure. Higher $\Delta$ indicate groupings which deviate from the cluster as a whole in either their mean velocity or dispersion (or both). For reference, $\Delta=309.9$ for our cluster sample of 224 galaxies. The significance is evaluated by Monte Carlo shuffles of the velocities while holding the positions fixed; this is the null hypothesis that there is no relationship between position and velocity. Graphically, each galaxy in the cluster may be represented by a circle with radius proportional to $e^{\delta_i}$. Thus, collections of larger circles indicate regions of substructure. Such regions are readily apparent in Figure \ref{fig-dstest}, particularly to the northeast and southwest (and also slightly to the north of the cluster center).

In an attempt to identify substructures and determine which galaxies compose these substructures, we applied the KMM algorithm \citep[see][]{ashm1994}. This algorithm divides a system into a specified number of components and calculates the Gaussian distance (using both positions and velocities) of each galaxy to each component.\footnote{The ``Gaussian distance'' for a given galaxy to component $i$ is defined as $f_i = \mbox{exp}[-((\alpha - \bar\alpha_i)/2\sigma_{\alpha_i})^2 -((\delta - \bar\delta_i)/2\sigma_{\delta_i})^2 -((v - \bar v_i)/2\sigma_{v_i})^2]$, where $\bar\alpha_i$ is the mean and $\sigma_{\alpha_i}$ is the dispersion in RA for galaxies in component $i$, etc. See \citet{mill2003} for additional details.} The galaxies are assigned to their nearest component, new means and dispersions for the positions and velocities of each component are calculated, and the procedure is repeated. The algorithm converges when the assignments are stable. In addition to the assignments, the probability that a given assignment is accurate may be assessed through the Gaussian distances of that galaxy to each component. For example, if a galaxy were equidistant to two components its assignment would only imply $50\%$ confidence. It should be stressed that the algorithm is statistical and does not necessarily imply association with a real physical substructure. Any number of components may be specified by the user and the algorithm will attempt to apportion the data into those components. In some cases, the algorithm will either not converge or will remove nearly all galaxies from a given component, indicating the specified component is not viable. 

Based on the observation that the velocity histogram had a long tail at higher velocities, our first KMM run divided the cluster into two components, a primary and a higher velocity subcluster to the southwest. Noting the multiple regions for substructure in the graphical representation of the $\Delta$ test (Figure \ref{fig-dstest}), we proceeded to a three-component fit. This fit kept the higher velocity subcluster to the southwest largely intact, and split off a lower velocity system to the northeast. This same general impression is obtained from an adaptively-smoothed distribution of the galaxy surface density, as shown in Figure 5 of \citet{owen1999}. Details of these KMM runs are provided in Tables \ref{tbl-kmm} and \ref{tbl-kmm2}, and the results of the three-component fit are depicted in Figures \ref{fig-dist} and \ref{fig-kmmvh}.

A slight digression on the validity of the KMM assignments is in order. The original astronomical application of KMM by \citet{ashm1994} was to detect bimodality in univariate datasets. For this purpose, and assuming the data are split into two components of equal dispersion but different mean, the authors found KMM to be relatively insensitive to input parameters. As might be expected, we have found through extensive testing that addition of variables largely removes this independence on input conditions. Clearly, adding two additional variables (for position) and relaxing the constraint that different components have common dispersions in a given variable expands the set of possible solutions. For example, with carefully chosen input parameters we were able to produce a two-component fit which largely ignored velocity and separated the cluster into a main component and a southwest clump, and a three-component fit which largely ignored position and sub-divided the higher velocity galaxies into two components. The merits of these various solutions are difficult to debate; \citet{ashm1994} note that the significance at which an additional substructure improves the fit to the data is difficult to determine in all but the fairly simple case of a single variable decomposed into two components of equal dispersion but different mean. It is largely for these reasons that we did not extend the KMM analysis to more than three components. We note here that other solutions exist, but the two we discuss were driven by the substructure test results and in particular by the locations of substructure indicated by the $\Delta$ test (Figure \ref{fig-dstest}).

The KMM assignments enable us to revisit Figure \ref{fig-dstest}, the Dressler-Shectman bubble plot (refer also to Figure \ref{fig-dist}). Three regions of possible substructure are identified from the plot as collections of larger circles: one to the northeast, one to the southwest, and one slightly north of the cluster core. These may be caused either by groupings of galaxies at a velocity differing from the cluster as a whole, or by groupings of galaxies with local dispersions differing from the cluster as a whole (refer to Equation \ref{eqn-ds}). The possible northeast substructure has larger $\delta$ values primarily because its galaxies are at a lower velocity than the cluster. This corresponds to the lower velocity component identified in the KMM runs fitting more than two substructures. The possible southwest substructure has larger $\delta$ values both because: 1) its galaxies have a higher local systemic velocity than the remainder of the cluster, and 2) the higher velocity galaxies are seen in projection with galaxies at the same general velocity as the cluster, thus increasing the velocity dispersion among nearest neighbors (in particular, note the concentration of higher velocity galaxies to the southeast of the main southwest clump). This region corresponds to the higher velocity component identified in the KMM runs. Finally, the higher local $\delta$ values just to the north of the cluster core are the result of large local velocity dispersion. This is caused by the superposition of galaxies from the low velocity substructure, the high velocity substructure, and the main body of the cluster.

\section{Discussion}\label{sec-discuss}

\subsection{The Dynamical State of A2125}

The above analysis indicates that A2125 possesses very strong evidence for substructure. What is the cause of this behavior? There are several possible interpretations. It is possible that we are simply viewing a projection of multiple clusters. The velocity separation of the main cluster and the higher velocity system implies a physical separation of $\sim$45 Mpc, assuming the difference is simply due to the Hubble flow. Their close alignment along the line of sight would then be fairly remarkable. \citet{bahc2003} used early Sloan Digital Sky Survey data to compile a cluster catalog out to $z\sim0.3$, and from these data determined an estimate of the space density of galaxy clusters. Should clusters be distributed randomly, this space density implies that the probability of any two clusters being seen at an angular separation of less than or equal to that implied by the KMM fits is 2.2$\%$. If we restrict this further to clusters at the redshifts implied by the KMM fits, the probability is only 0.1$\%$. Of course, this is a gross estimate; it ignores the reality of large-scale structure (which would increase the probability of such an alignment) but also does not take into account the richness of the clusters (the SDSS catalog includes substantially poorer clusters, so the chance of two clusters as rich as implied by the velocity dispersions in Table \ref{tbl-kmm} being aligned would be lower). In addition, a simple superposition does not entirely fit with the X-ray data. Figure \ref{fig-dist} includes X-ray contours derived from the smoothed {\it ROSAT} data. The cluster galaxies reported in this study match reasonably well with the X-ray features, with concentrations tracing the core of A2125 and the more diffuse X-ray emission to the southwest (labelled ``LSBXE'' after \citet{wang1997}). Although there is substantial overlap in the positional distributions of galaxies in the KMM fits, a rough mapping has the higher velocity system associated with the southwest X-ray emission. But as noted in \citet{wang1997} and \citet{wang2004}, this region is not simply explained as a single cluster or group: it is too diffuse to be a virialized cluster (and is composed primarily of spiral galaxies), and although its X-ray temperature ($\sim$1 keV) is representative of groups its luminosity (over $10^{43}$ \ergss) is in excess of typical group luminosities by an order of magnitude. Our velocity data also indicate a much higher dispersion than would be expected for a single poor cluster or group, and similarly that the velocities are poorly fit by a Gaussian (Figure \ref{fig-kmmvh}, although the Gaussian hypothesis for this component can not be rejected via either the $I$ or $KS$ tests). Thus, these X-ray papers favor that the diffuse X-ray emission to the southwest arises from hot intergalactic gas in a hierarchical structure including multiple groups. Presumably, this structure is oriented roughly along our line of sight and connects to the core of A2125.

Another possibility is that we are witnessing a cluster-cluster merger. This interpretation is based largely on comparison of the substructure test results with the N-body simulations of cluster-cluster mergers presented by PRBB. In general, the one-dimensional tests (i.e., those performed using only the galaxy velocities) are excellent indicators of mergers viewed near core passage when the relative velocity between the merging partners is greatest. Of course, this requires that the merger not be viewed orthogonal to the merger axis. Conversely, the 2D (positions only) tests perform best at early epochs of mergers and when the system is viewed orthogonal to the merger axis, thereby maximizing the positional separation of the merging partners. The 3D tests combine the strengths and weaknesses of their lower dimensional counterparts.

Such consistently strong indications of substructure as are found in A2125 (refer to Table \ref{tbl-MDsig}) are fairly rare in the simulations. Even so, the results are generally consistent with either of two possibilities. The first is the earlier stages of a merger (up to 2 Gyr prior to core passage) in which the positional separation of the two components is large, and consequently the system is viewed closer to orthogonal to the merger axis. The second possibility is a merger seen near the epoch when the cores of the respective clusters have become coincident and with a viewing angle more closely aligned with the merger axis (within about 30\degree). Since this latter possibility is the same as that obtained from the 1D substructure tests, this scenario best matches the A2125 data. It is also interesting to note that the two multi-dimensional tests which do not suggest overly strong evidence for a merger are the two least sensitive tests identified by the simulations. These are the Angular Separation Test \citep[AST;][]{west1988} and the $\epsilon$ test \citep{bird1993}. The AST looks for substructure in the form of small angular separations where the angle between any two galaxies is defined using the cluster centroid as the vertex. It therefore performs best when subclusters are well separated, a condition that fails around the time of core passage or if the system is viewed along the merger axis. The $\epsilon$ test calculates the projected mass estimator in the vicinity of galaxies using their nearest neighbors. Regions of substructure have lower projected mass estimates than the global mass estimate, and the significance of this deviation is tested via Monte Carlo shuffles of the data. Although the $\epsilon$ test is sensitive to mergers, it is most sensitive for times well after core passage (PRBB).

If the merger hypothesis is correct, whether the system is being viewed before or after core passage is dependent on the viability of the lower velocity component in the KMM fits. PRBB note that the difference in the velocity dispersions of the two components (for ease of reference, hereafter called the primary and the subcluster) is a good diagnostic of the timing of the merger (see their Table 9). Prior to core passage, the velocity dispersion of the primary is larger than that of the subcluster simply because of its greater mass. As the subcluster merges with the primary, it is dispersed and thereby obtains the larger dispersion. In the two-component KMM fit, the subcluster has a slightly lower velocity dispersion and would then indicates that the system is being viewed shortly before core passage (see Table \ref{tbl-kmm}). Removing the lower velocity galaxies from the primary when going to a three-component fit reduces the dispersion of the primary below that of the subcluster, suggesting the system is viewed shortly after core passage. In either case the high relative velocity of subcluster to the primary, $1940$ \kms{} in the two-component fit and $1702$ \kms{} in the three-component fit, is consistent with viewing a merger near the time of core passage \citep[e.g., see Equation 15 and associated discussion in][]{sara2001}.

A simple sanity check on the merger hypothesis is to ask whether the presumed merging parties are bound. Treating the primary and subcluster as point masses, the simple Newtonian condition for a bound system is:
\begin{equation}
V_r^2 R_p \leq 2 G M \mbox{sin}^2 \alpha ~\mbox{cos} ~\alpha
\end{equation}
where $V_r$ is the line of sight velocity difference between the two masses, $R_p$ is their projected separation, $M$ is the total mass, and $\alpha$ is the projection angle defined from the plane of the sky. The KMM analysis provides a reference from which to evaluate this condition, in particular by providing estimates for $V_r$, $R_p$, and $M$ (under the assumption that each the primary and the subcluster represent virialized systems). In the two-component fit, $V_r = 1940$ \kms, $R_p = 700$ kpc, and $M = 3.4 \times 10^{15}$ M$_\odot$ (the virial mass for the primary and subcluster are $2.1 \times 10^{15}$ and $1.3 \times 10^{15}$, respectively). Thus, the system is bound for nearly all projection angles ($6^\circ \leq \alpha \leq 84^\circ$). Only if the total mass of the system is very low ($\lesssim 8 \times 10^{14}$ M$_\odot$) would the system be unbound. Similarly, using the two largest components in the three-component fit the system would be bound for $5^\circ \leq \alpha \leq 85^\circ$ and unbound if $M \lesssim 5 \times 10^{14}$ M$_\odot$. Note that in this latter case, the virial assumption is surely suspect as the subcluster has the larger velocity dispersion and hence implied mass ($0.9 \times 10^{15}$ M$_\odot$ for the primary and $1.4 \times 10^{15}$ M$_\odot$ for the subcluster).

This merger hypothesis is potentially further supported by the X-ray morphology, although as with the case of a simple superposition there are difficulties. The X-ray emission of the main body of A2125 is elongated in a direction perpendicular to the low surface brightness X-ray emission. This elongation is consistent with ``an initial coalescence of subunits of comparable masses'' \citep{wang1997}, whereas at later stages the elongation would tend to be along the supercluster filament \citep[see also the simulations of ][]{roet1997}. However, it is clear that the X-ray emission from the main body of A2125 represents only a small portion of the entire system (refer to Figure \ref{fig-dist}). While a merger of some sort is indicated by the X-ray data, the scale of this merger is debatable.

In any case, we are observing A2125 under special circumstances which may largely explain some of the peculiarities of the cluster. At Richness Class 4, it is among the very richest clusters of galaxies known  \citep[e.g.,][]{morr2003}. Similarly, its large velocity dispersion suggests that it is a massive cluster containing many galaxies. However, these factors are inconsistent with the low X-ray luminosity and temperature of the cluster  \citep[$L_X = 7.9 \times 10^{43}$ ergs s$^{-1}$ and $T = 3.2$  keV;][]{wang2004}. For example, using the $\sigma-T$ relationship of  \citet{lubin93} the velocity dispersion of A2125 would predict a temperature of  7.5 keV. Such a discrepancy can be explained if the system is composed of multiple complexes seen in projection, whether these components are merging or not. A more appropriate application of the  $\sigma-T$ relationship would then be for an individual cluster in the system.  Working backwards, the X-ray temperature predicts a velocity dispersion of about  670 \kms, a figure in line with the velocity dispersions of the larger substructures (i.e., components with the most members) identified in the KMM analysis (refer to Table \ref{tbl-kmm}).

\subsection{Emission Line vs. Non-Emission Line Galaxies}

As prior studies \citep[e.g.,][]{owen1999,morr2003} have noted an excess in the fraction of active galaxies in A2125 relative to other clusters of comparable richness, we performed a simple investigation of activity as defined by presence of emission lines. For this purpose, the noted presence of any emission line in a galaxy's spectrum regardless of the strength of that line was sufficient for inclusion of that galaxy in the emission line sample. Using this definition, 75 of 224 cluster galaxies ($33.5\%$) were emission line objects. Unfortunately, we lack comparable data sets for other intermediate-redshift clusters to test whether this fraction is unusually high. For comparison, 96 of the 151 non-cluster galaxies in the field of A2125 ($63.6\%$) were emission line objects. 

When measured over the entire cluster, the dynamics of the emission line galaxies do not differ significantly from those of the non-emission line galaxies. The systemic velocity and dispersion for the emission line galaxies were $73798\pm133$ \kms{} and 1148$^{+107}_{-84}$ \kms, respectively, while those for the non-emission line galaxies were $73950\pm90$ \kms{} and $1092^{+70}_{-58}$ \kms. Similarly, there is no significant evidence that the emission line galaxies are drawn from a different population than the cluster galaxies as a whole, as quantified by KS and Wilcoxon tests. The positional data suggest that the emission line galaxies are at lower declinations, although this is significant only at 1.85$\sigma$. We also checked whether the emission line galaxies were preferentially associated with any substructures identified via the KMM analysis. A $\chi ^2$ test indicated that there was no statistical difference in the fraction of galaxies with emission lines among the various groups. Thus, the emission line galaxies are fairly well mixed throughout the cluster.

\section{Conclusions}\label{sec-conclude}

We have reported on a program which has obtained 371 galaxy velocities in the  direction of the $z=0.2465$ rich cluster Abell 2125. These new velocities include 222 cluster galaxies (including 25 in common with those reported in \citet{owen1999}) which increases the total number of confirmed  cluster members to 224, making this among the better studied intermediate redshift Butcher-Oemler clusters. The galaxy velocities were subjected to a large number of statistical tests which identified significant substructure. Through comparison with numerical simulations, these tests indicated that A2125 is consistent with a merger of comparable-mass clusters viewed along an axis within $\sim30$\degree{} of the merger axis. This merger is viewed within a few hundred million years of the time at which the cores of the respective components have crossed. 

A favorable alignment of multiple structures, be they involved in a merger or simply seen in projection, is helpful in understanding the general properties of A2125. The combination of these structures produces an unusually high richness and velocity dispersion, reconciling the apparent discrepancy of the richness of the cluster with its low X-ray temperature and luminosity.

\acknowledgments
We are deeply saddened by the passing of Michael Ledlow. His enthusiasm for astronomy was surpassed only by the kindness of his heart, and he will be sorely missed.

The authors thank Jason Pinkney, who kindly provided his fortran code which performed the various substructure tests. We also thank Sam Barden at NOAO for his excellent advice and support of the Hydra multifiber spectrograph, and an anonymous referee for useful comments which improved the paper. NAM acknowledges the support of a National Research Council Associateship award held at NASA's Goddard Space Flight Center.

\clearpage

\begin{deluxetable}{l c r c}
\tablecolumns{4}
\tablecaption{Summary of Observations\label{tbl-obs}}
\tablewidth{0pt}
\tablehead{
\colhead{Field ID} & \colhead{Date} & \colhead{Exposures} & \colhead{Instrument}
}
\startdata
H1   & 1996 April 16 & $6 \times 1800$ & WIYN+Hydra \\
H2   & 1996 April 16 & $2 \times 2400$ & WIYN+Hydra \\
     & 1996 April 17 & $4 \times 2400$ & WIYN+Hydra \\
H3   & 1996 April 17 & $3 \times 2580$ & WIYN+Hydra \\
     & 1996 April 18 & $4 \times 2580$ & WIYN+Hydra \\
H4   & 1996 May 20   & $4 \times 2100$ & WIYN+Hydra \\
     & 1996 May 21   & $2 \times 2100$ & WIYN+Hydra \\
     & 1996 May 22   & $2 \times 2100$ & WIYN+Hydra \\
\tableline
1187 & 1998 May 27   & $2 \times 1800$ & Mayall multislits \\
1188 & 1998 May 27   & $2 \times 1800$ & Mayall multislits \\
1189 & 1998 May 27   & $2 \times 1800$ & Mayall multislits \\
1191 & 1998 May 27   & $2 \times 1800$ & Mayall multislits \\
1196 & 1998 May 28   & $2 \times 1800$ & Mayall multislits \\
1197 & 1998 May 28   & $2 \times 1800$ & Mayall multislits \\
1198 & 1998 May 28   & $2 \times 1800$ & Mayall multislits \\
1199 & 1998 May 28   & $2 \times 1800$ & Mayall multislits \\
1200 & 1998 May 28   & $2 \times 1800$ & Mayall multislits \\
1190 & 1998 May 29   & $2 \times 1800$ & Mayall multislits \\
1192 & 1998 May 29   & $2 \times 1800$ & Mayall multislits \\
1193 & 1998 May 29   & $2 \times 1800$ & Mayall multislits \\
1194 & 1998 May 29   & $2 \times 1800$ & Mayall multislits \\
\tableline
Q101 & 2000 March 30 & $4 \times 2800$ & WIYN+Hydra (queue) \\
     & 2000 April 2  & $2 \times 2800$ & WIYN+Hydra (queue) \\
Q102 & 2000 May 27   & $6 \times 2800$ & WIYN+Hydra (queue) \\
Q103 & 2000 May 28   & $5 \times 2800$ & WIYN+Hydra (queue) \\
\enddata

\tablecomments{Exposure times are in seconds.}
\end{deluxetable}

\begin{deluxetable}{l l r r c c}
\tablecolumns{6}
\tablecaption{Cluster Velocity Data\label{tbl-vels}}
\tablewidth{444pt}
\tablehead{
\colhead{RA} & \colhead{Dec} & \colhead{$cz$} & \colhead{Error} &
\colhead{Lines} & \colhead{Source}
}
\startdata
15 38 13.9 & +66 13 06 &  74199 &  81 & \nodata                     & W00 \\
15 39 02.0 & +66 05 24 &  77196 &  81 & \nodata                     & W00 \\
15 39 08.8 & +66 08 54 &  73449 &  49 & \nodata                     & W00 \\
15 39 22.2 & +66 12 42 &  73419 &  40 & \OII, \OIII                 & W00 \\
15 39 23.1 & +66 19 58 &  73068 &  40 & \OII, \Hb                   & W96 \\
15 39 23.5 & +66 11 53 &  73226 &  81 & \nodata                     & W96 \\
15 39 27.3 & +66 10 44 &  76660 &  67 & \nodata                     & W96 \\
15 39 36.4 & +66 12 19 &  73676 &  88 & (\OII)                      & W96 \\
15 39 36.8 & +66 12 27 &  73387 &  59 & \nodata                     & W96 \\*
           &           &  73359 &  51 & \nodata                     & W00 \\
15 39 37.2 & +66 07 10 &  72738 &  63 & \nodata                     & W96 \\
15 39 37.3 & +66 17 44 &  73047 &  51 & \nodata                     & W96 \\
15 39 40.1 & +66 04 29 &  73989 &  40 & \OII, \Hb, \OIII, \Ha, \NII & W00 \\
15 39 40.3 & +66 14 46 &  74431 &  66 & \nodata                     & W96 \\ 
15 39 41.5 & +66 11 13 &  72672 &  54 & \nodata                     & W96 \\
15 39 45.4 & +66 10 35 &  72602 &  63 & \nodata                     & W96 \\*
           &           &  72460 &  49 & \nodata                     & M98 \\
15 39 45.7 & +66 08 06 &  74791 &  70 & \nodata                     & W96 \\
15 39 46.4 & +66 10 22 &  72622 &  57 & \nodata                     & W96 \\*
           &           &  72700 &  57 & \nodata                     & M98 \\
15 39 46.9 & +66 07 34 &  75098 &  57 & \nodata                     & W96 \\ 
15 39 48.0 & +66 11 05 &  72248 &  52 & \nodata                     & W96 \\*
           &           &  72250 &  49 & \nodata                     & M98 \\
15 39 49.2 & +66 22 12 &  73871 &  40 & \OII, \Hb, \OIII            & W96 \\
15 39 49.7 & +66 07 09 &  73446 &  40 & \Ha, \NII                   & W00 \\
15 39 50.1 & +66 14 55 &  73621 &  61 & \nodata                     & W96 \\
15 39 50.5 & +66 05 13 &  73189 &  40 & \OII, \Hb                   & W96 \\
15 39 50.5 & +66 10 13 &  72666 &  59 & \nodata                     & W96 \\
15 39 51.6 & +66 10 44 &  72190 &  40 & \Hb, \Ha, \NII              & M98 \\
15 39 52.5 & +66 13 19 &  74408 &  40 & \Ha, \NII                   & W00 \\
15 39 52.6 & +66 09 54 &  73648 &  54 & \nodata                     & W96 \\*
           &           &  73689 &  46 & \nodata                     & W00 \\
15 39 53.0\tablenotemark{a} & +66 10 48 &  73546 &  51 & \nodata    & W96 \\*
           &           &  73749 &  57 & \nodata    & M98 \\*
           &           &  73389 &  57 & \nodata                     & M98 \\
15 39 55.6 & +66 18 30 &  73492 &  52 & \nodata                     & W96 \\
15 39 56.2 & +66 16 32 &  72787 &  59 & \nodata                     & W96 \\
15 39 56.4 & +66 17 04 &  73908 &  62 & \nodata                     & W96 \\
15 39 57.2 & +66 07 36 &  73493 &  62 & \nodata                     & W96 \\
15 39 59.3 & +66 11 27 &  73796 &  40 & \OII, \Hb, \OIII            & W96 \\*
           &           &  73734 &  40 & \OII, \Hb, \OIII, \Ha, \NII & M98 \\
15 39 59.4 & +66 16 08 &  73684 &  55 & \nodata                     & W96 \\
15 40 00.3 & +66 10 00 &  73479 &  61 & \nodata                     & M98 \\
15 40 02.0 & +66 09 53 &  72640 &  40 & \OII, \Ha, \NII             & M98 \\
15 40 02.2 & +66 14 18 &  73809 &  40 & \Ha, \NII                   & W00 \\
15 40 04.3 & +66 18 16 &  74662 &  54 & \nodata                     & W96 \\
15 40 04.7 & +66 09 31 &  73929 &  61 & \nodata                     & M98 \\
15 40 05.2 & +66 10 53 &  75638 &  57 & \nodata                     & M98 \\
15 40 05.3 & +66 10 56 &  75818 &  52 & \nodata                     & W96 \\*
           &           &  75757 &  51 & \nodata                     & M98 \\
15 40 05.4 & +66 10 13 &  72619 &  51 & \nodata                     & W96 \\*
           &           &  72700 &  49 & \nodata                     & W00 \\*
           &           &  72550 &  47 & \nodata                     & M98 \\*
15 40 06.6 & +66 10 13 &  73029 &  40 & \OII, \Hb, \OIII, \Ha, \NII & W00 \\
15 40 07.5 & +66 12 26 &  74105 &  55 & \nodata                     & W96 \\*
           &           &  74049 &  49 & \nodata                     & M98 \\
15 40 07.9 & +66 13 57 &  77091 &  62 & \nodata                     & W96 \\
15 40 09.0 & +66 17 59 &  73760 &  62 & \nodata                     & W96 \\
15 40 09.0 & +66 12 17 &  73585 &  53 & \nodata                     & W96 \\*
           &           &  73509 &  49 & \nodata                     & M98 \\
15 40 09.8 & +66 09 52 &  73734 &  78 & \nodata                     & W96 \\*
           &           &  73999 &  57 & \nodata                     & M98 \\
15 40 10.3 & +66 11 29 &  76890 &  50 & \nodata                     & W96 \\*
           &           &  76657 &  51 & \nodata                     & M98 \\
15 40 10.4 & +66 15 46 &  74170 &  55 & \nodata                     & W96 \\
15 40 11.5 & +66 19 09 &  77075 &  60 & \nodata                     & W96 \\
15 40 11.8 & +66 18 59 &  71811 &  53 & \nodata                     & W96 \\
15 40 12.0 & +66 12 10 &  76700 &  51 & \nodata                     & W96 \\*
           &           &  76627 &  49 & \nodata                     & W00 \\
15 40 13.1 & +66 10 01 &  73749 &  40 & \OII, \Hb, \OIII, \Ha       & M98 \\
15 40 13.6 & +66 18 24 &  74746 &  49 & \nodata                     & W96 \\
15 40 13.9 & +66 11 50 &  73750 &  40 & \OII, \Hb, \OIII            & W96 \\* 
           &           &  73689 &  40 & \OII, \Hb, \OIII, \Ha, \NII & W00 \\
15 40 14.8 & +66 08 44 &  75849 &  74 & \nodata                     & W96 \\
15 40 15.0 & +66 17 06 &  75334 &  66 & \nodata                     & W96 \\
15 40 15.4 & +66 18 03 &  72904 &  40 & \OII, \Hb, \OIII            & W96 \\*
           &           &  72939 &  40 & \OII, \Hg, \Hb, \OIII, \Ha, \NII & W00 \\
15 40 15.8 & +66 16 17 &  75301 &  52 & \nodata                     & W96 \\
15 40 15.8 & +66 14 07 &  74588 &  61 & \nodata                     & W00 \\
15 40 15.9 & +66 11 10 &  73591 &  40 & \OII, \Hb, \OIII            & W96 \\*
           &           &  73839 &  40 & \Hb, \OIII, \Ha, \NII       & M98 \\
15 40 16.5 & +66 10 40 &  76619 &  51 & \nodata                     & W96 \\*
           &           &  76537 &  51 & \nodata                     & M98 \\
15 40 16.5 & +66 12 43 &  73002 &  48 & \nodata                     & W96 \\*
           &           &  72640 &  61 & \nodata                     & M98 \\
15 40 17.1 & +66 11 16 &  73255 &  70 & (\OII, \Hb)                 & W96 \\*
           &           &  73209 &  40 & \OII, \Ha                   & M98 \\
15 40 18.1 & +66 12 18 &  74558 &  61 & \nodata                     & M98 \\
15 40 19.2 & +66 15 03 &  73722 &  51 & \nodata                     & W96 \\
15 40 20.6 & +66 27 29 &  73253 &  72 & \nodata                     & W96 \\
15 40 21.3 & +66 09 04 &  73752 &  60 & \nodata                     & W96 \\
15 40 21.4 & +66 10 12 &  76813 &  49 & \nodata                     & W96 \\*
           &           &  76897 &  46 & \nodata                     & W00 \\
15 40 22.3 & +66 27 07 &  74449 &  76 & \nodata                     & W96 \\
15 40 23.1 & +66 25 00 &  74044 &  56 & \nodata                     & W96 \\
15 40 23.2 & +66 08 58 &  73921 &  51 & \nodata                     & W96 \\
15 40 23.8 & +66 11 53 &  72999 &  57 & \nodata                     & M98 \\
15 40 23.9 & +66 16 21 &  74954 &  51 & \nodata                     & W96 \\
15 40 24.7 & +66 14 30 &  73149 &  69 & \nodata                     & W00 \\
15 40 28.0 & +66 15 02 &  74199 &  40 & \OII, \Hb, \OIII, \Ha, \NII & W00 \\
15 40 28.4 & +66 04 17 &  75487 &  65 & \nodata                     & W96 \\
15 40 29.4 & +66 06 55 &  72491 &  50 & \nodata                     & W96 \\*
           &           &  72580 &  47 & \nodata                     & W00 \\
15 40 30.1 & +66 12 14 &  73578 &  40 & \OII, \Hb                   & W96 \\* 
           &           &  73689 &  40 & \OII, \Hb, \OIII, \Ha, \NII & M98 \\
15 40 30.3 & +66 13 04 &  74758 &  49 & \nodata                     & W96 \\*
           &           &  74888 &  61 & \nodata                     & M98 \\
15 40 30.9 & +66 12 26 &  74169 &  40 & \OII, \Ha                   & M98 \\
15 40 31.3 & +66 12 31 &  74198 &  40 & \OII, \OIII                 & W96 \\
15 40 31.3 & +66 17 29 &  72790 &  56 & \nodata                     & W96 \\*
           &           &  72819 &  49 & \nodata                     & W00 \\
15 40 31.5 & +66 15 13 &  73269 &  40 & \OII, \Hb, \OIII, \Ha       & W00 \\
15 40 33.0 & +66 08 47 &  72794 &  54 & (\OII)                      & W96 \\* 
           &           &  72879 &  49 & \nodata                     & W00 \\
15 40 33.3 & +66 17 06 &  76255 &  51 & \nodata                     & W96 \\*
           &           &  76447 &  49 & \nodata                     & W00 \\
15 40 36.4 & +66 12 28 &  72640 &  40 & \Hb, \Ha, \NII              & W00 \\
15 40 39.1 & +66 06 48 &  76541 &  40 & \OII, \Hb, \OIII            & W96 \\
15 40 39.1 & +66 24 06 &  71380 &  40 & \OII, \Hb, \OIII, \Ha, \NII & W00 \\
15 40 39.3 & +66 17 35 &  74542 &  74 & \nodata                     & W96 \\
15 40 40.1 & +66 13 09 &  73724 &  40 & \OII, \Hb, \OIII            & W96 \\*
           &           &  73718 &  40 & \OII, \Hg, \Hb, \OIII, \Ha, \NII, \SII & M98 \\
15 40 41.5 & +66 19 57 &  72884 &  73 & \nodata                     & W96 \\*
           &           &  72819 &  40 & \Hb, \Ha, \NII              & W00 \\
15 40 42.4 & +66 04 42 &  75560 &  40 & \OII, \Hb                   & W96 \\*
           &           &  75578 &  40 & \OII, \Ha, \NII             & W00 \\
15 40 43.2 & +66 10 21 &  76687 &  40 & \OII, \Hb, \Ha, \NII        & W00 \\
15 40 44.3 & +66 18 46 &  73059 &  57 & \nodata                     & W00 \\*
           &           &  73419 &  81 & \nodata                     & M98 \\
15 40 47.0 & +66 07 31 &  76841 &  40 & \OII, \Hb                   & W96 \\ 
15 40 47.0 & +66 09 34 &  75216 &  84 & (\OII, \Hb)                 & W96 \\ 
15 40 47.2 & +66 18 38 &  73079 &  52 & \nodata                     & W96 \\*
           &           &  73149 &  49 & \nodata                     & M98 \\
15 40 47.3 & +66 19 06 &  71974 &  51 & \nodata                     & W96 \\*
           &           &  72340 &  61 & \nodata                     & M98 \\
15 40 48.5 & +66 13 02 &  72640 &  53 & \nodata                     & W00 \\
15 40 48.6 & +66 05 25 &  76761 &  52 & \nodata                     & W96 \\
15 40 49.0 & +66 18 36 &  73299 &  61 & \nodata                     & M98 \\
15 40 49.2 & +66 18 40 &  72972 &  54 & \nodata                     & W96 \\*
           &           &  73059 &  57 & \nodata                     & M98 \\
15 40 51.8 & +66 10 39 &  75337 &  56 & \nodata                     & W96 \\
15 40 52.0 & +66 04 55 &  75092 &  73 & (\OII, \Hb)                 & W96 \\
15 40 52.2 & +66 14 28 &  72518 &  40 & \OII, \Hb, \OIII            & W96 \\* 
           &           &  72670 &  40 & \OII, \OIII, \Ha, \NII      & M98 \\
15 40 52.2 & +66 25 20 &  74409 &  59 & \nodata                     & W96 \\
15 40 52.7 & +66 12 31 &  77551 &  40 & \OII, \Hb                   & W96 \\*
           &           &  77586 &  40 & \OII, \Hb, \Ha              & W00 \\
15 40 53.6 & +66 05 26 &  75147 &  66 & (\OII)                      & W96 \\*
           &           &  75188 &  40 & \OII, \Ha, \NII             & W00 \\
15 40 53.7 & +66 16 07 &  76297 &  61 & \nodata                     & W00 \\
15 40 54.5 & +66 11 28 &  74468 &  40 & \OII, \Hb, \Ha, \NII        & M98 \\
15 40 54.6 & +66 17 57 &  72727 &  52 & \nodata                     & W96 \\*
           &           &  72819 &  51 & \nodata                     & M98 \\
15 40 54.7 & +66 17 16 &  76819 &  58 & \nodata                     & W96 \\*
           &           &  76777 &  40 & \OII, \Hb, \Ha, \NII        & W00 \\*
           &           &  76747 &  40 & \OII, \Ha, \NII             & M98 \\
15 40 55.6 & +66 16 45 &  74528 &  61 & \nodata                     & M98 \\
15 40 55.9 & +66 18 39 &  72280 &  61 & \nodata                     & M98 \\ 
15 40 56.9 & +66 26 46 &  73787 &  62 & \nodata                     & W96 \\*
           &           &  73929 &  49 & \nodata                     & W00 \\
15 40 57.0 & +66 16 46 &  74217 &  58 & \nodata                     & W96 \\
15 40 57.0 & +66 26 39 &  73329 &  67 & \nodata                     & W96 \\
15 40 57.2 & +66 26 10 &  73437 &  57 & \nodata                     & W96 \\
15 40 57.3 & +66 15 22 &  76297 &  40 & \Hb, \Ha                    & W00 \\
15 40 57.4 & +66 17 03 &  73620 &  51 & \nodata                     & W96 \\*
           &           &  73749 &  49 & \nodata                     & M98 \\
15 41  0.3 & +66 19 03 &  71686 &  51 & \nodata                     & W96 \\*
           &           &  71800 &  45 & \nodata                     & W00 \\
15 41 01.7 & +66 16 33 &  74079 &  57 & \nodata                     & M98 \\
15 41 01.9 & +66 16 27 &  73679 &  49 & \nodata                     & W96 \\*
           &           &  73659 &  46 & \nodata                     & W00 \\*
           &           &  73629 &  51 & \nodata                     & M98 \\
15 41 02.0 & +66 17 21 &  76479 &  70 & \nodata                     & W96 \\*
           &           &  76957 &  69 & \nodata                     & M98 \\
15 41 02.3 & +66 20 51 &  72062 &  57 & \nodata                     & W96 \\
15 41 02.4 & +66 18 30 &  73569 &  40 & \OII, \Hb, \OIII, \Ha, \NII & W00 \\*
           &           &  73749 &  40 & \OII, \Ha                   & M98 \\
15 41 02.6 & +66 16 37 &  75098 &  69 & \nodata                     & W00 \\
15 41 03.2 & +66 13 05 &  73037 &  56 & \nodata                     & W96 \\
15 41 04.7 & +66 15 51 &  74225 &  65 & \nodata                     & W96 \\*
           &           &  74199 &  61 & \nodata                     & M98 \\
15 41 05.0 & +66 18 23 &  77047 &  40 & \OII, \Hb                   & M98 \\
15 41 05.4 & +66 16 15 &  73269 &  57 & \nodata                     & M98 \\
15 41 05.4 & +66 13 07 &  72640 &  40 & \Hb, \OIII, \Ha             & W00 \\
15 41 05.7 & +66 15 09 &  73197 &  40 & \OII, \Hb, \OIII            & W96 \\* 
           &           &  73029 &  40 & \OII, \Hb, \OIII, \Ha       & M98 \\
15 41 06.2 & +66 15 17 &  73959 &  69 & \nodata                     & W00 \\
15 41 06.9 & +66 19 49 &  72103 &  70 & (\OII)                      & W96 \\
15 41 07.0 & +66 18 03 &  71770 &  81 & \nodata                     & M98 \\
15 41 07.4 & +66 15 59 &  75098 &  61 & \nodata                     & W00 \\
15 41 09.8 & +66 15 45 &  75846 &  58 & (\OII, \Hb, \OIII)          & W96\tablenotemark{b} \\*
           &           &  75697 &  40 & \OII, \OIII, \Ha, \NII      & M98 \\
15 41 09.9 & +66 15 25 &  74768 &  61 & \nodata                     & M98 \\
15 41 10.5 & +66 11 16 &  73496 &  59 & \nodata                     & W96 \\
15 41 10.9 & +66 17 21 &  74376 &  54 & \nodata                     & W96 \\*
           &           &  74408 &  61 & \nodata                     & M98 \\
15 41 11.0 & +66 26 01 &  73843 &  89 & \nodata                     & W96 \\
15 41 11.7 & +66 12 48 &  74139 &  40 & \OII, \Hb, \Ha, \NII        & W00 \\
15 41 13.3 & +66 20 56 &  71936 &  70 & \nodata                     & W96 \\ 
15 41 14.1 & +66 14 55 &  73082 &  52 & \nodata                     & W96 \\*
           &           &  73119 &  46 & \nodata                     & W00 \\
15 41 14.3 & +66 17 36 &  75993 &  50 & \nodata                     & W96 \\*
           &           &  76027 &  51 & \nodata                     & M98 \\
15 41 14.4 & +66 15 57 &  75389 &  50 & \nodata                     & W96 \\*
           &           &  75368 &  47 & \nodata                     & W00 \\*
           &           &  75518 &  46 & \nodata                     & M98 \\
15 41 14.7 & +66 26 28 &  74061 &  50 & \nodata                     & W96 \\
15 41 14.9 & +66 16 04 &  73971 &  47 & \nodata                     & W96 \\*
           &           &  73929 &  46 & \nodata                     & M98 \\
15 41 15.4 & +66 15 59 &  73929 &  49 & \nodata                     & W00 \\
15 41 16.1 & +66 26 00 &  73175 &  52 & \nodata                     & W96 \\
15 41 16.6 & +66 17 43 &  72630 &  66 & (\OII, \Hb)                 & W96 \\*
           &           &  72610 &  40 & \OII, \Hb, \Ha, \NII        & M98 \\
15 41 17.3 & +66 19 24 &  73899 &  51 & \nodata                     & W00 \\
15 41 17.3 & +66 21 16 &  72154 &  54 & \nodata                     & W96 \\
15 41 17.6 & +66 16 33 &  72536 &  53 & \nodata                     & W96 \\*
           &           &  72370 &  61 & \nodata                     & M98 \\
15 41 17.7 & +66 12 30 &  72408 &  40 & \OII, \Hb, \OIII            & W96 \\
15 41 18.3 & +66 15 47 &  75727 &  61 & \nodata                     & M98 \\
15 41 18.6 & +66 11 25 &  74140 &  40 & \OII, \Hb                   & W96 \\
15 41 19.0 & +66 15 24 &  74348 &  49 & \nodata                     & W00 \\
15 41 19.3 & +66 16 31 &  75836 &  52 & \nodata                     & W96 \\*
           &           &  75697 &  51 & \nodata                     & M98 \\
15 41 19.9 & +66 26 53 &  71511 &  58 & \nodata                     & W96 \\
15 41 20.2 & +66 26 15 &  73434 &  51 & \nodata                     & W96 \\
15 41 22.4 & +66 15 04 &  74502 &  73 & \nodata                     & W96 \\
15 41 22.7 & +66 17 34 &  73899 &  61 & \nodata                     & M98 \\
15 41 22.8 & +66 17 01 &  72216 &  40 & \OII, \Hb                   & W96 \\*
           &           &  72190 &  40 & \OII, \Hb, \OIII            & M98 \\
15 41 23.0 & +66 17 04 &  74434 &  84 & \nodata                     & W96 \\
15 41 24.2 & +66 24 41 &  70742 &  57 & \nodata                     & W96 \\
15 41 24.2 & +66 16 05 &  73538 &  68 & \nodata                     & W96 \\*
           &           &  73599 &  47 & \nodata                     & W00 \\
15 41 27.6 & +66 08 45 &  73536 &  50 & \nodata                     & W96 \\
15 41 28.5 & +66 21 20 &  74468 &  40 & \OII, \Hb, \OIII, \Ha, \NII & W00 \\ 
15 41 30.3 & +66 10 37 &  74841 &  40 & \OII, \Hb, \OIII            & W96 \\
15 41 30.5 & +66 14 55 &  75518 &  69 & \nodata                     & W00 \\
15 41 30.8 & +66 17 40 &  75428 &  40 & \OII, \Hb, \Ha, \NII        & W00 \\
15 41 31.5 & +66 21 31 &  76300 &  53 & \nodata                     & W96 \\
15 41 32.6 & +66 14 26 &  73729 &  52 & \nodata                     & W96 \\
15 41 33.2 & +66 17 42 &  75188 &  40 & \OII, \Ha                   & W00 \\
15 41 33.9 & +66 22 55 &  75368 &  61 & \nodata                     & W00 \\
15 41 33.9 & +66 31 12 &  71021 &  61 & \nodata                     & W00 \\
15 41 34.0 & +66 19 59 &  72613 &  52 & \nodata                     & W96 \\
15 41 34.4 & +66 14 23 &  73459 &  52 & \nodata                     & W96 \\
15 41 35.1 & +66 14 03 &  73689 &  69 & \nodata                     & W00 \\
15 41 35.9 & +66 16 15 &  76027 &  61 & \nodata                     & W00 \\
15 41 37.3 & +66 22 50 &  74588 &  40 & \OII, \Hb, \Ha, \NII        & W00 \\
15 41 38.6 & +66 20 32 &  74889 &  67 & \nodata                     & W96 \\
15 41 38.9 & +66 17 13 &  71860 &  40 & \OII, \Hb, \Ha, \NII        & W00 \\
15 41 40.1 & +66 18 34 &  71436 &  60 & \nodata                     & W96 \\
15 41 40.1 & +66 13 41 &  76177 &  40 & \OII, \Hb, \Ha, \NII        & W00 \\
15 41 40.3 & +66 24 46 &  73763 &  53 & \nodata                     & W96 \\
15 41 41.2 & +66 27 12 &  71870 &  40 & \OII, \Hb                   & W96 \\
15 41 41.5 & +66 10 46 &  75458 &  57 & \nodata                     & W00 \\
15 41 42.8 & +66 16 41 &  72123 &  52 & \nodata                     & W96 \\
15 41 43.2 & +66 15 17 &  75383 &  58 & \nodata                     & W96 \\
15 41 43.4 & +66 14 19 &  76447 &  62 & \nodata                     & W96 \\
15 41 45.5 & +66 17 07 &  73537 &  57 & (\OII)                      & W96 \\
15 41 47.6 & +66 17 57 &  74281 &  52 & \nodata                     & W96 \\*
           &           &  74348 &  53 & \nodata                     & W00 \\
15 41 48.5 & +66 13 58 &  73307 &  55 & \nodata                     & W96 \\
15 41 48.6 & +66 25 46 &  74708 &  40 & \OII, \Ha, \NII             & W00 \\
15 41 49.1 & +66 12 18 &  74139 &  40 & \Ha, \NII                   & W00 \\
15 41 50.6 & +66 26 28 &  70606 &  62 & \nodata                     & W96 \\
15 41 51.3 & +66 18 16 &  74749 &  59 & \nodata                     & W96 \\
15 41 52.7 & +66 21 48 &  74503 &  40 & \OII, \Hb, \OIII            & W96 \\
15 42 02.8 & +66 15 55 &  72049 &  66 & (\OII)                      & W96 \\*
           &           &  72310 &  40 & \Ha, \NII                   & W00 \\
15 42 03.3 & +66 26 49 &  74059 &  78 & (\OII)                      & W96 \\*
           &           &  73989 &  40 & \Hb, \Ha, \NII              & W00 \\
15 42 03.9 & +66 26 32 &  73680 &  40 & \OII, \OIII                 & W96 \\
15 42 07.1 & +66 27 15 &  73815 &  60 & \nodata                     & W96 \\
15 42 12.1 & +66 19 07 &  74998 &  40 & \OII, \Hb                   & W96 \\*
           &           &  74948 &  40 & \OII, \Hb, \Ha              & W00 \\
15 42 24.3 & +66 19 58 &  74109 &  53 & \nodata                     & W00 \\
           &           &  74139 &  61 & \nodata                     & M98 \\
15 42 26.0 & +66 21 12 &  71614 &  49 & \nodata                     & W96 \\
15 42 26.1 & +66 07 18 &  72206 &  40 & \OII, \OIII                 & W96 \\
15 42 32.9 & +66 07 00 &  72275 &  40 & \OII, \Hb, \OIII            & W96 \\
15 42 34.8 & +66 18 15 &  74373 &  63 & \nodata                     & W96 \\
15 42 38.4 & +66 18 22 &  74643 &  40 & \OII, \Hb, \OIII            & W96 \\
15 42 49.7 & +66 21 34 &  71752 &  51 & (\OII)                      & W96 \\
15 42 51.8 & +66 12 34 &  73874 &  56 & \nodata                     & W96 \\
15 42 56.4 & +66 14 07 &  74369 &  66 & \nodata                     & W96 \\
15 42 56.7 & +66 18 07 &  74515 &  50 & \nodata                     & W96 \\
15 43 01.9 & +66 23 25 &  71473 &  40 & \OII, \Hb, \OIII            & W96 \\
15 43 07.8 & +66 13 42 &  71856 &  40 & \OII, \Hb, \OIII            & W96 \\* 
           &           &  71770 &  40 & \Hb, \OIII, \Ha, \NII       & W00 \\
15 43 11.1 & +66 11 35 &  77295 &  40 & \OII, \Hb                   & W96 \\
\enddata

\tablenotetext{a}{Binary galaxy, with each nucleus observed individually in the long slit observations. The 1996 WIYN Hydra observation is likely a blend of the two nuclei.}

\tablenotetext{b}{This is the peculiar emission line galaxy C153, discussed in more detail in \citet{dwar2001}.}

\tablecomments{All coordinates are J2000, and heliocentric velocities (note that these are not relativistically corrected, and are therefore $cz$) and errors are in \kms. Emission lines used in determination of the reported velocities are indicated. When these are enclosed in parentheses, the velocity was determined through cross correlation but the noted emission lines were also detected at a consistent velocity. The abbreviations for the data sources are: W96 = WIYN Hydra 1996, W00 = WIYN Hydra 2000, and M98 = Mayall 1998.}

\end{deluxetable}


\begin{deluxetable}{l l c c c c}
\tablecolumns{6}
\tablecaption{Non-cluster Redshifts\label{tbl-xvels}}
\tablewidth{444pt}
\tablehead{
\colhead{RA} & \colhead{Dec} & \colhead{$z$} & \colhead{Error} &
\colhead{Lines} & \colhead{Source}
}
\startdata
15 37 56.1 & +66 27 36 & 0.2049 & 0.0003 & \nodata                     & W00 \\
15 38 12.4 & +66 14 05 & 0.4373 & 0.0001 & \OII, \OIII                 & W00 \\
15 39 00.4 & +65 59 48 & 0.2878 & 0.0002 & \nodata                     & W00 \\
15 39 20.1 & +66 14 28 & 0.3365 & 0.0003 & \nodata                     & W00 \\
15 39 20.1 & +65 57 23 & 1.9010 & 0.0010 & C~{\scshape iv}, O~{\scshape iii}], C~{\scshape iii}] & W00 \\
15 39 22.5 & +66 18 26 & 2.2140 & 0.0010 & C~{\scshape iv}, C~{\scshape iii}] & W00 \\
15 39 22.7 & +66 20 12 & 0.0231 & 0.0001 & \Hb, \OIII, \Ha, \SII       & W96 \\
15 39 32.5 & +66 24 01 & 0.1210 & 0.0002 & \nodata                     & W96 \\
15 39 33.0 & +66 05 48 & 0.0320 & 0.0001 & \Hb, \OIII, \Ha             & W96 \\
15 39 33.1 & +66 07 44 & 0.4808 & 0.0002 & \nodata                     & W00 \\
15 39 34.0 & +66 05 26 & 0.0323 & 0.0001 & \Hb, \OIII, \Ha, \NII, \SII & W96 \\
15 39 35.7 & +66 25 18 & 0.3822 & 0.0001 & \OII, \Hb, \OIII            & W96 \\*
           &           & 0.3820 & 0.0001 & \OII, \Hg, \Hb, \OIII, \Ha, \NII & W00 \\
15 39 35.8 & +66 11 27 & 0.3612 & 0.0001 & \OII, \Hb, \OIII            & W00 \\
15 39 36.1 & +66 21 41 & 0.6776 & 0.0001 & \OII, \Hb                   & W00 \\
15 39 36.9 & +66 07 31 & 0.3224 & 0.0001 & \nodata                     & W96 \\
15 39 37.9 & +66 21 03 & 0.4375 & 0.0001 & \OII, \Hb, \OIII            & W00 \\
15 39 40.5 & +66 24 33 & 0.1222 & 0.0002 & \nodata                     & W96 \\
15 39 41.5 & +66 05 03 & 0.1228 & 0.0001 & \Hb, \OIII, \Ha             & W96 \\
15 39 42.5 & +66 24 56 & 0.4078 & 0.0002 & \nodata                     & W00 \\
15 39 45.4 & +66 12 37 & \tablenotemark{a} & \tablenotemark{a} & \nodata                     & W00 \\
15 39 46.4 & +66 21 25 & 0.5226 & 0.0001 & \OII, \Hb, \OIII            & W00 \\
15 39 48.8 & +65 57 43 & 0.1225 & 0.0001 & \Hb, \OIII, \Ha, \SII       & W00 \\
15 39 53.3 & +66 18 32 & 0.3718 & 0.0003 & \nodata                     & W00 \\
15 39 53.3 & +66 10 42 & 0.6389 & 0.0001 & \OII, \OIII                 & W00 \\
15 39 54.3 & +66 27 20 & 0.2031 & 0.0001 & \OII, \Hb, \OIII            & W96 \\
15 39 56.1 & +66 05 33 & 0.2755 & 0.0001 & \OII, \Hb                   & W96 \\*
           &           & 0.2759 & 0.0001 & \OII, \Hb                   & W00 \\
15 39 58.4 & +66 13 10 & 0.3411 & 0.0002 & \nodata                     & W96 \\
15 40 00.2 & +66 18 58 & 0.2877 & 0.0001 & \OII, \OIII                 & W00 \\
15 40 03.4 & +66 22 22 & 0.6760 & 0.0001 & \OII, \Hb                   & W00 \\
15 40 04.8 & +66 13 49 & 0.5465 & 0.0003 & \nodata                     & W00 \\
15 40 07.3 & +66 35 46 & 0.2949 & 0.0002 & \nodata                     & W00 \\
15 40 07.7 & +66 15 36 & 0.0930 & 0.0001 & \Hb, \OIII, \Ha             & W96 \\
15 40 08.5 & +66 12 26 & 0.3223 & 0.0003 & \nodata                     & W96 \\
15 40 11.3 & +66 23 07 & 0.3067 & 0.0003 & \nodata                     & W00 \\
15 40 25.4 & +66 31 01 & 0.4401 & 0.0002 & \nodata                     & W00 \\
15 40 26.6 & +66 15 30 & 0.3794 & 0.0001 & \OII, \Hb, \Ha              & W00 \\
15 40 29.2 & +66 10 47 & 0.3894 & 0.0001 & \OII, \Hg, \Hb, \OIII       & M98 \\
15 40 29.6 & +66 24 42 & 0.0807 & 0.0001 & \Hb, \OIII, \Ha             & W96 \\
15 40 30.8 & +66 22 24 & 0.3413 & 0.0001 & \OII, \OIII                 & W96 \\
15 40 32.0 & +66 20 33 & 0.3622 & 0.0001 & \OII, \OIII                 & W96 \\
15 40 33.6 & +66 08 02 & 0.9183 & 0.0001 & Mg~{\scshape ii}, \OII      & W00 \\
15 40 33.8 & +66 07 25 & 0.2752 & 0.0001 & \OII, \Hb                   & W00 \\
15 40 35.3 & +66 12 05 & 0.2692 & 0.0001 & \nodata                     & W00 \\
15 40 35.3 & +66 24 56 & 0.0808 & 0.0001 & \OIII, \Ha                  & W96 \\ 
15 40 37.7 & +66 08 24 & 0.6248 & 0.0002 & \nodata                     & W00 \\
15 40 40.8 & +66 26 54 & 0.0563 & 0.0001 & \Ha, \NII, \SII             & W96 \\
15 40 42.6 & +66 08 54 & 0.6264 & 0.0003 & \nodata                     & W00 \\
15 40 44.9 & +66 09 05 & \tablenotemark{a} & \tablenotemark{a} & \nodata                     & W00 \\ 
15 40 48.3 & +66 15 17 & 0.3231 & 0.0001 & \OII, \OIII                 & W96 \\
15 40 49.5 & +66 09 38 & 0.1964 & 0.0001 & \OII, \OIII                 & W96 \\
15 40 49.5 & +66 15 41 & 0.1965 & 0.0003 & \nodata                     & M98 \\
15 40 49.6 & +66 21 06 & 0.0683 & 0.0003 & \nodata                     & W00 \\
15 40 50.2 & +66 25 53 & 0.9143 & 0.0010 & Mg~{\scshape ii}            & W96 \\
15 40 50.5 & +66 13 00 & 0.2204 & 0.0003 & \nodata                     & W00 \\
15 40 51.0 & +66 16 32 & 0.1270 & 0.0002 & (\Ha, \NII)                 & W96 \\
           &           & 0.1270 & 0.0001 & \Ha, \NII, \SII             & M98 \\
15 40 51.7 & +66 23 43 & 0.0376 & 0.0001 & \Hb, \OIII, \Ha, \SII       & W96 \\
15 40 51.9 & +66 06 31 & 0.3632 & 0.0003 & (\OII)                      & W96 \\*
           &           & 0.3657 & 0.0002 & \nodata                     & W00 \\
15 40 52.7 & +66 19 14 & 0.1057 & 0.0001 & \Hb, \OIII, \Ha, \NII, \SII & W96 \\*
           &           & 0.1059 & 0.0001 & \OII, \Hb, \OIII, \Ha, \NII, \SII & M98 \\
15 40 54.3 & +66 21 32 & 0.2957 & 0.0001 & \OII, \Hg, \Hb, \OIII, \Ha, \NII, \SII & W00 \\
15 40 54.6 & +66 06 38 & 0.3605 & 0.0003 & \nodata                     & W96 \\
15 40 56.4 & +66 11 38 & 0.0426 & 0.0001 & \Hb, \OIII, \Ha, \NII, \SII & W96 \\
15 40 56.4 & +66 16 28 & 1.0120 & 0.0010 & Mg~{\scshape ii}, \OII      & W00 \\
15 41 00.8 & +66 13 54 & 0.3614 & 0.0003 & \nodata                     & W00 \\
15 41 03.6 & +65 58 11 & 0.1244 & 0.0001 & \Ha, \NII, \SII             & W00 \\
15 41 05.3 & +66 12 36 & 0.2749 & 0.0001 & \OII, \Hb, \OIII            & W96 \\* 
           &           & 0.2752 & 0.0003 & (\OII)                      & M98 \\
15 41 06.3 & +66 20 20 & 0.1094 & 0.0001 & \Hb, \OIII, \Ha, \NII, \SII & W96 \\
15 41 07.6 & +66 16 57 & 0.2969 & 0.0003 & \nodata                     & W00 \\
15 41 10.6 & +66 13 14 & 0.2270 & 0.0003 & \nodata                     & M98 \\
15 41 11.1 & +66 14 13 & 0.0935 & 0.0001 & \Hb, \OIII                  & W96 \\ 
15 41 11.5 & +66 21 11 & 0.0863 & 0.0001 & \Hb, \OIII, \Ha, \NII       & W96 \\
15 41 11.6 & +66 14 08 & 0.0932 & 0.0001 & \Hb, \OIII, \Ha, \NII, \SII & W96 \\*
           &           & 0.0933 & 0.0001 & \Hb, \OIII, \Ha, \NII, \SII & W00 \\
15 41 12.5 & +66 17 17 & 0.2704 & 0.0003 & \nodata                     & W00 \\
15 41 12.9 & +66 15 02 & 0.5068 & 0.0003 & \nodata                     & W00 \\
15 41 14.6 & +66 21 40 & 0.0855 & 0.0001 & \Ha, \NII                   & W96 \\*
           &           & 0.0854 & 0.0001 & \Hg, \Hb, \OIII, \Ha, \NII, \SII & W00 \\
15 41 17.0 & +66 16 27 & 0.6921 & 0.0003 & (\OII)                      & W00 \\ 
15 41 17.3 & +66 07 06 & 0.3217 & 0.0003 & \nodata                     & W00 \\
15 41 17.8 & +66 19 10 & 0.3344 & 0.0003 & \nodata                     & M98 \\
15 41 17.9 & +66 18 49 & 0.3366 & 0.0003 & \nodata                     & M98 \\
15 41 20.9 & +66 19 21 & 0.3597 & 0.0001 & \OII, \Hb                   & W00 \\
15 41 22.1 & +65 54 39 & 0.0783 & 0.0002 & \nodata                     & W00 \\
15 41 24.3 & +66 07 57 & 0.3627 & 0.0001 & \OII, \Hb                   & W00 \\
15 41 24.7 & +66 25 02 & 0.3398 & 0.0001 & \OII, \Hb, \OIII            & W96 \\
15 41 28.1 & +66 13 25 & 0.0931 & 0.0001 & \Hb, \OIII, \Ha, \NII, \SII & W96 \\
15 41 28.6 & +66 11 53 & 0.1265 & 0.0001 & \Hb, \OIII, \Ha             & W96 \\
15 41 31.5 & +66 22 06 & 0.3395 & 0.0003 & \nodata                     & W96 \\
15 41 33.3 & +66 08 33 & 0.3368 & 0.0003 & \nodata                     & W00 \\
15 41 35.1 & +66 16 01 & 0.2752 & 0.0002 & \nodata                     & W96 \\
15 41 36.1 & +66 18 50 & 0.3404 & 0.0001 & \OII, \Hb, \OIII            & W00 \\
15 41 37.9 & +66 13 10 & 0.5229 & 0.0001 & \OII, \Hg, \Hb, \OIII       & W00 \\
15 41 41.0 & +66 22 38 & 1.3820 & 0.0010 & C~{\scshape iii}], Mg~{\scshape ii} & W00 \\
15 41 44.7 & +66 09 15 & 0.1131 & 0.0002 & \nodata                     & W96 \\
15 41 44.9 & +66 13 49 & 0.2113 & 0.0003 & \nodata                     & W00 \\
15 41 46.3 & +66 11 16 & 0.5274 & 0.0002 & \nodata                     & W00 \\
15 41 47.3 & +66 17 08 & 0.3220 & 0.0003 & \nodata                     & W00 \\
15 41 47.4 & +66 11 25 & 0.5262 & 0.0003 & \nodata                     & W00  \\
15 41 48.9 & +66 11 37 & 0.5263 & 0.0002 & \nodata                     & W00 \\
15 41 52.9 & +66 08 52 & 0.1266 & 0.0001 & \Hb, \OII, \Ha              & W96 \\
15 41 57.4 & +66 12 56 & 0.2070 & 0.0001 & \OII, \Hb, \OIII            & W96 \\*
           &           & 0.2070 & 0.0001 & \OII, \Hg, \Hb, \OIII, \Ha, \NII, \SII & W00 \\
15 42 00.5 & +66 13 36 & 0.1074 & 0.0001 & \Hb, \OII, \Ha              & W96 \\
15 42 01.1 & +66 10 12 & 0.5140 & 0.0002 & \nodata                     & W00 \\
15 42 01.5 & +66 11 40 & 0.1459 & 0.0002 & \Hb, \OIII                  & W96 \\
15 42 01.9 & +66 23 30 & 0.4284 & 0.0001 & \OII, \Hg, \Hb, \OIII\tablenotemark{b} & W00 \\
15 42 03.0 & +66 13 49 & 0.1482 & 0.0003 & (\OIII)                     & W96 \\ 
15 42 07.3 & +66 17 44 & 0.3437 & 0.0001 & \OII, \Hb, \OIII            & W00 \\
15 42 09.8 & +66 14 52 & 0.3596 & 0.0001 & \OII, \OIII                 & W00 \\
15 42 10.8 & +66 10 44 & 0.1091 & 0.0001 & \Hb, \OIII, \Ha             & W96 \\
15 42 11.4 & +66 07 15 & 0.1266 & 0.0001 & \Hb, \OIII, \Ha             & W96 \\
15 42 11.4 & +66 04 36 & 0.2248 & 0.0001 & \OII, \Hb, \OIII            & W96 \\* 
           &           & 0.2248 & 0.0001 & \OII, \Hb, \OIII, \Ha, \NII & W00 \\
15 42 14.6 & +66 30 04 & 0.2995 & 0.0002 & (\OII)                      & W00 \\
15 42 14.7 & +66 17 08 & 0.5112 & 0.0001 & \OII, \Hg, \Hb, \OIII       & W00 \\
15 42 17.5 & +66 27 26 & 0.1805 & 0.0001 & \OII, \Hb, \OIII            & W96 \\
15 42 18.5 & +66 25 56 & 0.2993 & 0.0001 & \OII, \OIII                 & W96 \\
15 42 18.9 & +66 15 27 & 0.1463 & 0.0002 & \nodata                     & W96 \\
15 42 21.9 & +66 12 24 & 0.2702 & 0.0001 & \OII, \Hb, \OIII            & W96 \\
15 42 21.9 & +66 21 08 & 0.3329 & 0.0001 & \OII, \Hb, \OIII            & W00 \\
15 42 23.7 & +66 18 36 & 0.3374 & 0.0002 & \nodata                     & W00 \\
15 42 24.6 & +66 21 20 & 0.2972 & 0.0002 & \nodata                     & W96 \\
15 42 24.7 & +66 10 12 & 0.2207 & 0.0001 & \OII, \Hb                   & W96 \\*
           &           & 0.2208 & 0.0001 & \OIII, \Ha, \NII, \SII      & W00 \\
15 42 25.1 & +66 09 17 & 0.6643 & 0.0001 & \OII, \Hb, \OIII            & W00 \\
15 42 29.8 & +66 14 28 & 0.5652 & 0.0001 & \OII, \OIII                 & W00 \\
15 42 31.6 & +66 14 26 & 0.0928 & 0.0001 & \Hb, \OIII, \Ha, \NII, \SII & W96 \\*
           &           & 0.0930 & 0.0001 & \Hb, \OIII, \Ha, \NII, \SII & W00 \\ 
15 42 39.2 & +66 13 27 & 0.3607 & 0.0003 & (\OII)                      & W96 \\*
           &           & 0.3612 & 0.0003 & \nodata                     & W00 \\
15 42 40.6 & +66 24 12 & 0.2755 & 0.0001 & \OII, \Hb, \OIII            & W96 \\*
           &           & 0.2754 & 0.0001 & \OII, \Hb, \OIII, \Ha, \NII, \SII & W00 \\
15 42 41.1 & +66 24 37 & 0.2758 & 0.0001 & \OII, \Hb, \OIII            & W96 \\
15 42 41.5 & +66 24 34 & 0.2751 & 0.0001 & \OII, \Hg, \Hb, \OIII, \Ha, \NII, \SII\tablenotemark{b} & W00 \\
15 42 42.1 & +66 20 36 & 0.6557 & 0.0001 & \OII, \Hb, \OIII            & W00 \\
15 42 42.4 & +66 09 40 & 0.0443 & 0.0001 & \Hb, \OIII, \Ha, \SII       & W96 \\
15 42 44.9 & +66 19 39 & 0.3319 & 0.0002 & \nodata                     & W96 \\
15 42 49.1 & +66 25 52 & 0.1469 & 0.0002 & \nodata                     & W96 \\
15 42 53.8 & +66 33 23 & 0.3239 & 0.0001 & \OII, \Hb, \OIII, \Ha       & W00 \\
15 42 54.4 & +66 18 09 & 0.5082 & 0.0002 & (\OII)                      & W00 \\
15 43 01.1 & +66 15 17 & 0.5113 & 0.0002 & \nodata                     & W00 \\
15 43 02.2 & +65 57 34 & 0.2767 & 0.0002 & \nodata                     & W00 \\
15 43 08.4 & +66 21 54 & 0.1246 & 0.0001 & \Hb, \OIII, \Ha, \NII       & W96 \\
15 43 09.3 & +66 17 30 & 0.2972 & 0.0003 & \nodata                     & W96 \\
15 43 09.8 & +66 26 20 & 0.2955 & 0.0001 & \OII, \Hb                   & W96 \\
15 43 10.5 & +66 17 57 & 0.5076 & 0.0002 & (\OII)                      & W00 \\
15 43 11.3 & +66 33 38 & 0.5704 & 0.0001 & \OII, \OIII                 & W00 \\
15 43 13.8 & +66 07 59 & 0.3319 & 0.0002 & \nodata                     & W96 \\*
           &           & 0.3319 & 0.0002 & \nodata                     & W00 \\
15 43 21.3 & +66 11 04 & 0.4346 & 0.0001 & \OII, \Hb                   & W00 \\
15 43 27.0 & +66 00 42 & 0.2016 & 0.0001 & \OIII, \Ha, \NII            & W00 \\
15 43 33.5 & +66 12 00 & 0.1130 & 0.0001 & \Hb, \Ha, \NII, \SII        & W00 \\
15 43 36.5 & +66 30 11 & 2.3099 & 0.0010 & C~{\scshape iv}, C~{\scshape iii}]? & W00 \\
15 43 42.2 & +66 13 38 & 0.3165 & 0.0001 & \OII, \OIII, \Ha            & W00 \\
15 43 43.1 & +66 09 28 & 0.7336 & 0.0001 & \OII, \OIII                 & W00 \\
15 43 44.9 & +66 30 00 & 0.3121 & 0.0003 & \nodata                     & W00 \\
15 43 46.3 & +66 07 20 & 0.6678 & 0.0001 & \OII, \Hg, \Hb, \OIII       & W00 \\
15 43 59.0 & +66 00 34 & 0.1498 & 0.0001 & \Hb, \OIII, \Ha, \NII, \SII & W00 \\
15 44 27.7 & +66 20 32 & 0.3118 & 0.0001 & \OII, \OIII                 & W00 \\
\enddata

\tablenotetext{a}{Probable quasar, velocity and error not listed as unique identification of emission features was not determined.}

\tablenotetext{b}{Strong narrow emission line galaxy. Other detected lines include [Ne~{\scshape iii}] \lam 3869, [He~{\scshape i}] + H \lam 3889, [Ne~{\scshape iii}] + H \lam 3967, H$\delta$ \lam 4102, and \OIII{} \lam 4363.}
 
\tablecomments{All coordinates are J2000, and reported redshifts and errors are heliocentric. Emission lines used in determination of the reported redshifts are indicated. When these are enclosed in parentheses, the redshift was determined through cross correlation but the noted emission lines were also detected at a consistent redshift. The abbreviations for the data sources are: W96 = WIYN Hydra 1996, W00 = WIYN Hydra 2000, and M98 = Mayall 1998.}

\end{deluxetable}

\begin{deluxetable}{c c c}
\tablecolumns{3}
\tablecaption{Normality Tests for A2125\label{tbl-1Dsig}}
\tablewidth{415pt}
\tablehead{
\colhead{Test} & \colhead{Statistic} & \colhead{Significance} 
}
\startdata
$A$           & 0.784 & \tablenotemark{a} \\
$U$           & 4.840 & 0.000 \\
$W$           & 0.965 & 0.001 \\
$B_1$         & 0.334 & 0.045\tablenotemark{b} \\
$B_2$         & 2.762 & 0.324\tablenotemark{b} \\
$B_1B_2$ Omni & 3.343 & 0.184\tablenotemark{b} \\
$I$           & 0.980 & 0.150\tablenotemark{c} \\
$KS$          & 1.017 & 0.025\tablenotemark{d} \\
$V$           & 1.714 & 0.010 \\
$W^2$         & 0.214 & 0.004 \\
$U^2$         & 0.185 & 0.005 \\
$A^2$         & 1.342 & 0.002 \\
$DIP$         & 0.017 & 0.030 \\
$AI$          & 0.109 & \tablenotemark{a} \\
$TI$          & 1.252 & \tablenotemark{a} \\
\enddata

\tablenotetext{a}{The significance for these tests can not be determined analytically. For comparison purposes, the $AI$ test statistic for a Gaussian is equal to 0.0 and for $TI$ it is equal to 1.0. See \citet{bird1993b} for more information on the $AI$ and $TI$ tests.}

\tablenotetext{b}{Based on the results for forty $N=150$ data sets drawn at random from the full $N=224$ set of velocities (see text). The dispersion in these values (i.e., the biweight scales) were: 0.041 for $B_1$, 0.122 for $B_2$, and 0.135 for the $B_1B_2$ Omni test. In each case, the test statistic derived from the forty $N=150$ data sets was nearly identical to that derived for the full $N=224$ data.}

\tablenotetext{c}{This is the default value for a non-significant deviation for the implementation of this test \citep[see][]{pink1996}. Note that all forty $N=150$ randomly-drawn subsets of the full velocity list yielded significance values of 0.15, indicating no significant deviation from a Gaussian. The $I$ test is normally evaluated and compared to $I_{90}$ to determine whether a distribution is non-Gaussian; distributions are considered consistent with a Gaussian if $I < I_{90}$. For the A2125 velocities, $I_{90} = 1.024$.}

\tablenotetext{d}{The significance values for this implementation of the KS test are restricted to 0.25, 0.15, 0.10, 0.05, 0.025, and 0.01. Using the forty randomly-drawn $N=150$ subsets, the robust mean significance is 0.055 with a disperson of 0.086.}

\tablecomments{The Significance represents the probability that the distribution is consistent with a Gaussian. Thus, a low number indicates likely non-Gaussian behavior.}

\end{deluxetable}

\begin{deluxetable}{l c c l}
\tablecolumns{4}
\tablecaption{Multidimensional Substructure Tests for A2125\label{tbl-MDsig}}
\tablewidth{0pt}
\tablehead{
\colhead{Test} & \colhead{Dimensionality} &
\colhead{Significance} & \colhead{Reference}
}
\startdata
Fourier Elongation & 2D & 0.000 & \citet{pink1996} \\
$\beta$            & 2D & 0.000 & \citet{west1988} \\
Angular Separation & 2D & 0.063 & \citet{west1988} \\
Lee                & 2D & 0.001 & \citet{fitc1988} \\
Lee                & 3D & 0.001 & \citet{pink1996} \\
$\Delta$           & 3D & 0.003 & \citet{dres1988} \\
$\epsilon$         & 3D & 0.130 & \citet{bird1993} \\
$\alpha$           & 3D & 0.000 & \citet{west1990} \\
$\alpha$ variant   & 3D & 0.000 & \citet{pink1996} \\
\enddata

\tablecomments{Implementations of all tests are described in \citet{pink1996}, and generally follow the prescriptions for the tests as set forth by the original authors. One minor exception is the number of nearest neighbors used, which was frequently 10 in the original tests (e.g., the $\Delta$ and $\alpha$ tests) but set to $\sqrt{N}$ in the \citet{pink1996} code (hence, $\sqrt{224}=14.97$ and the tests used the 15 nearest neighbors). Significance values are determined from 1000 Monte Carlo simulations.}

\end{deluxetable}

\clearpage

\begin{deluxetable}{c c c c c c c c c c c c}
\tablecolumns{12}
\tablecaption{Possible Substructures from KMM Runs: Velocity\label{tbl-kmm}}
\tabletypesize{\footnotesize}
\tablewidth{0pt}
\tablehead{
\colhead{Fitted} & \multicolumn{3}{c}{Component 1} & \colhead{} &
\multicolumn{3}{c}{Component 2} & \colhead{} & \multicolumn{3}{c}{Component 3} \\
\colhead{Components} & 
\colhead{$N$} & \colhead{$C_{BI}$} & \colhead{$S_{BI}$} & \colhead{} &
\colhead{$N$} & \colhead{$C_{BI}$} & \colhead{$S_{BI}$} & \colhead{} &
\colhead{$N$} & \colhead{$C_{BI}$} & \colhead{$S_{BI}$}
}
\startdata
1 & 224 & 73897 & 1113 & & \nodata & \nodata & \nodata & & \nodata & \nodata & \nodata \\
2 & 170 & 73425 & \phn712 & & 54 & 75970 & 603 & & \nodata & \nodata & \nodata \\
3 & 136 & 73656 & \phn479 & & 58 & 75889 & 623 & & 30 & 71926 & 362 \\
\enddata

\tablecomments{$C_{BI}$ and $S_{BI}$ refer to the robust estimates of the mean and dispersion for each fitted component, with values in units of \kms.}

\end{deluxetable}

\begin{deluxetable}{c c c c c c c c c}
\tablecolumns{9}
\tablecaption{Possible Substructures from KMM Runs: Position\label{tbl-kmm2}}
\tabletypesize{\footnotesize}
\tablewidth{0pt}
\tablehead{
\colhead{Fitted} & \multicolumn{2}{c}{Component 1} & \colhead{} &
\multicolumn{2}{c}{Component 2} & \colhead{} & \multicolumn{2}{c}{Component 3} \\
\colhead{Components} & 
\colhead{$RA(J2000)$} & \colhead{$Dec(J2000)$} & \colhead{} &
\colhead{$RA(J2000)$} & \colhead{$Dec(J2000)$} & \colhead{} &
\colhead{$RA(J2000)$} & \colhead{$Dec(J2000)$} 
}
\startdata
1 & 15:40:52 & 66:15:39 & & \nodata & \nodata & & \nodata & \nodata \\
2 & 15:40:53 & 66:16:23 & & 15:40:50 & 66:13:18 & & \nodata & \nodata \\
3 & 15:40:43 & 66:15:30 & & 15:40:53 & 66:13:39 & & 15:41:35 & 66:19:27 \\
\enddata

\tablecomments{Coordinates presented refer to the mean centers for each component. The errors in these means range from about one to several arcminutes.}

\end{deluxetable}

\begin{onecolumn}

\begin{figure}
\figurenum{1}
\epsscale{0.9}
\plotone{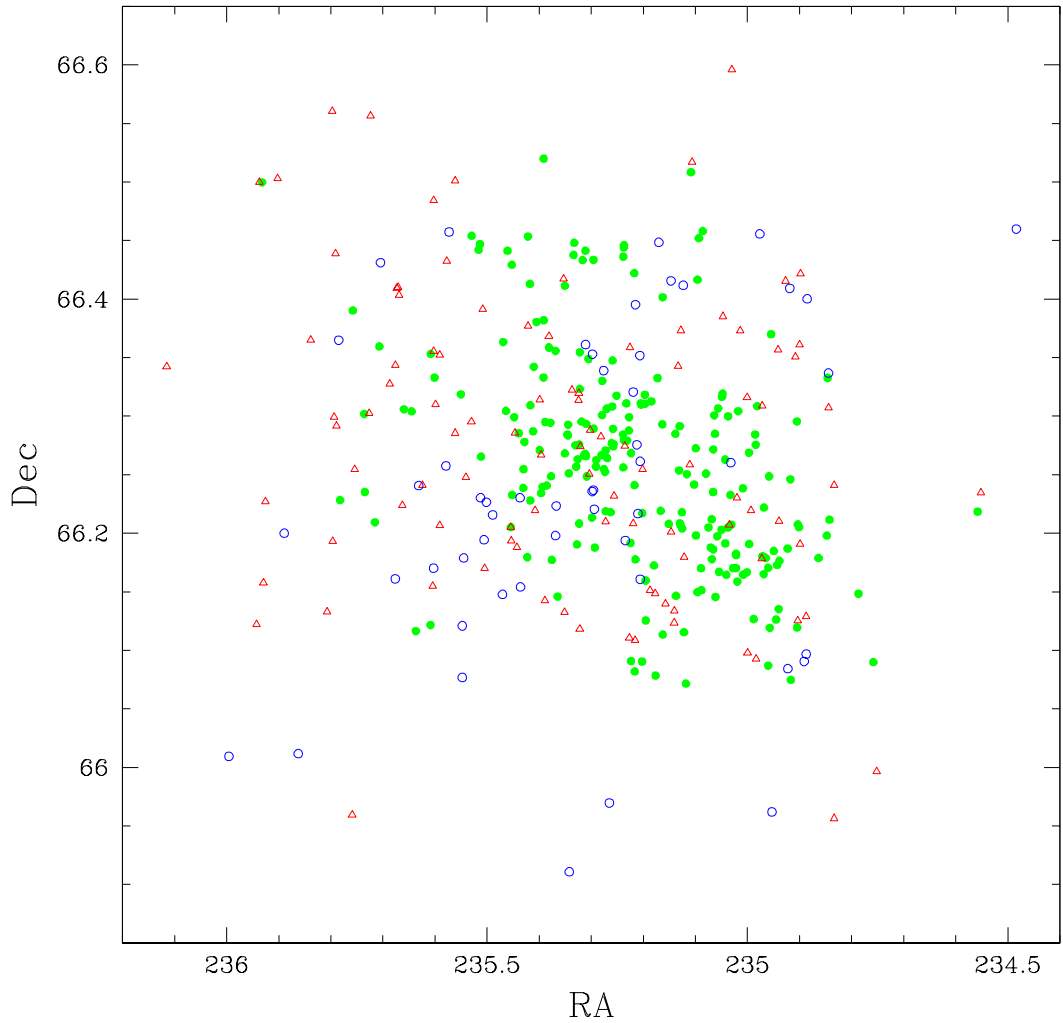}
\caption{Coded positions of all galaxies in the spectroscopic database. Cluster members are represented by filled green circles, background galaxies by open red triangles, and foreground galaxies by open blue circles. The relative uniformity of sampling is indicated by the fairly even distribution of background galaxies.\label{fig-allpos}}
\end{figure}

\begin{figure}
\figurenum{2}
\epsscale{0.9}
\plotone{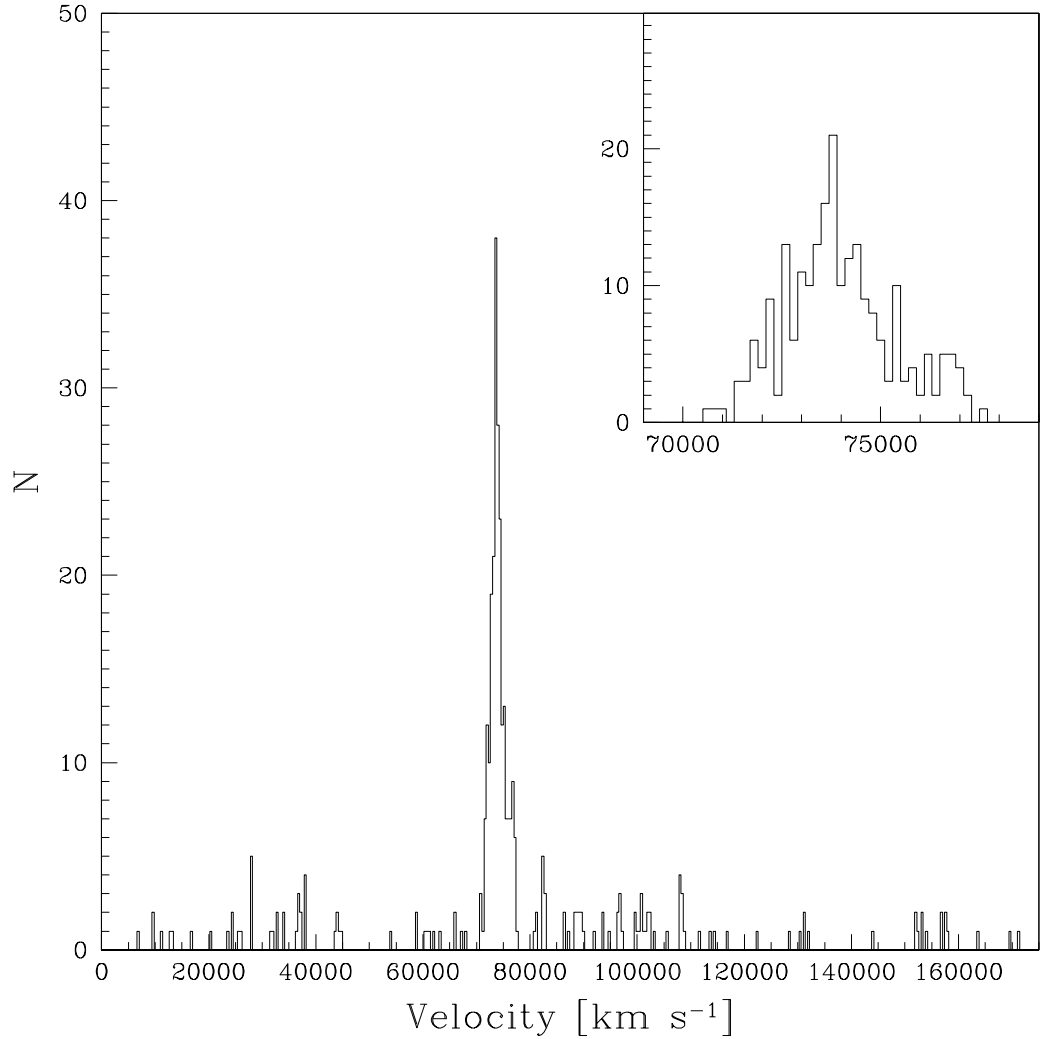}
\caption{Histogram for all data with $cz < 175000$ \kms, using a bin size of 400 \kms. The location of A2125 is easily apparent, and is detailed in the inset histogram. The binning for the inset is 200 \kms.\label{fig-vhfull}}
\end{figure}

\begin{figure}
\figurenum{3}
\epsscale{1.0}
\plotone{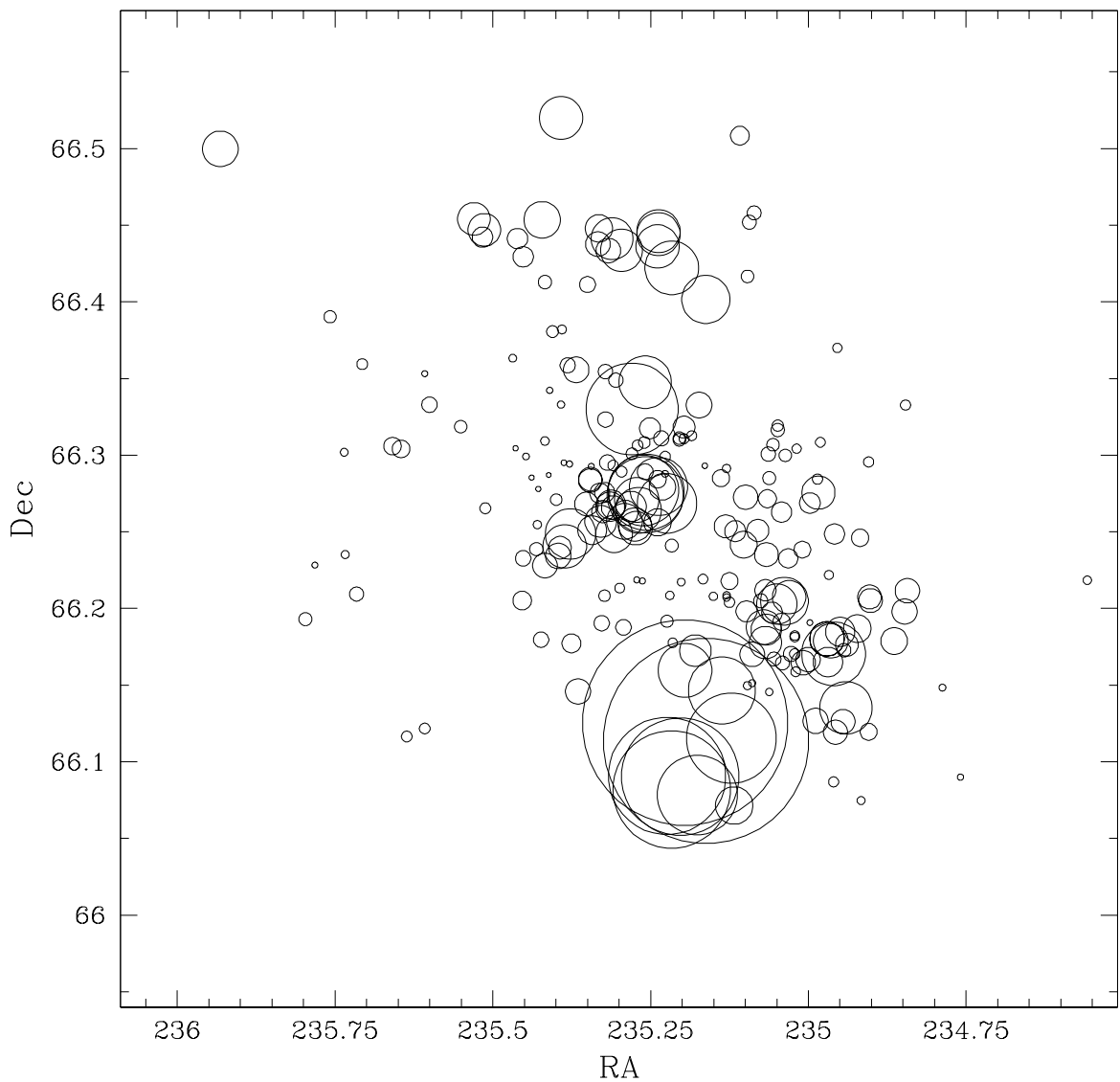}
\caption{Dressler-Shectman plot for A2125. Each galaxy is plotted as a circle with the diameter of the circle proportional to $e^\delta$.\label{fig-dstest}}
\end{figure}

\begin{figure}
\figurenum{4}
\epsscale{0.85}
\rotatebox{270}{\plotone{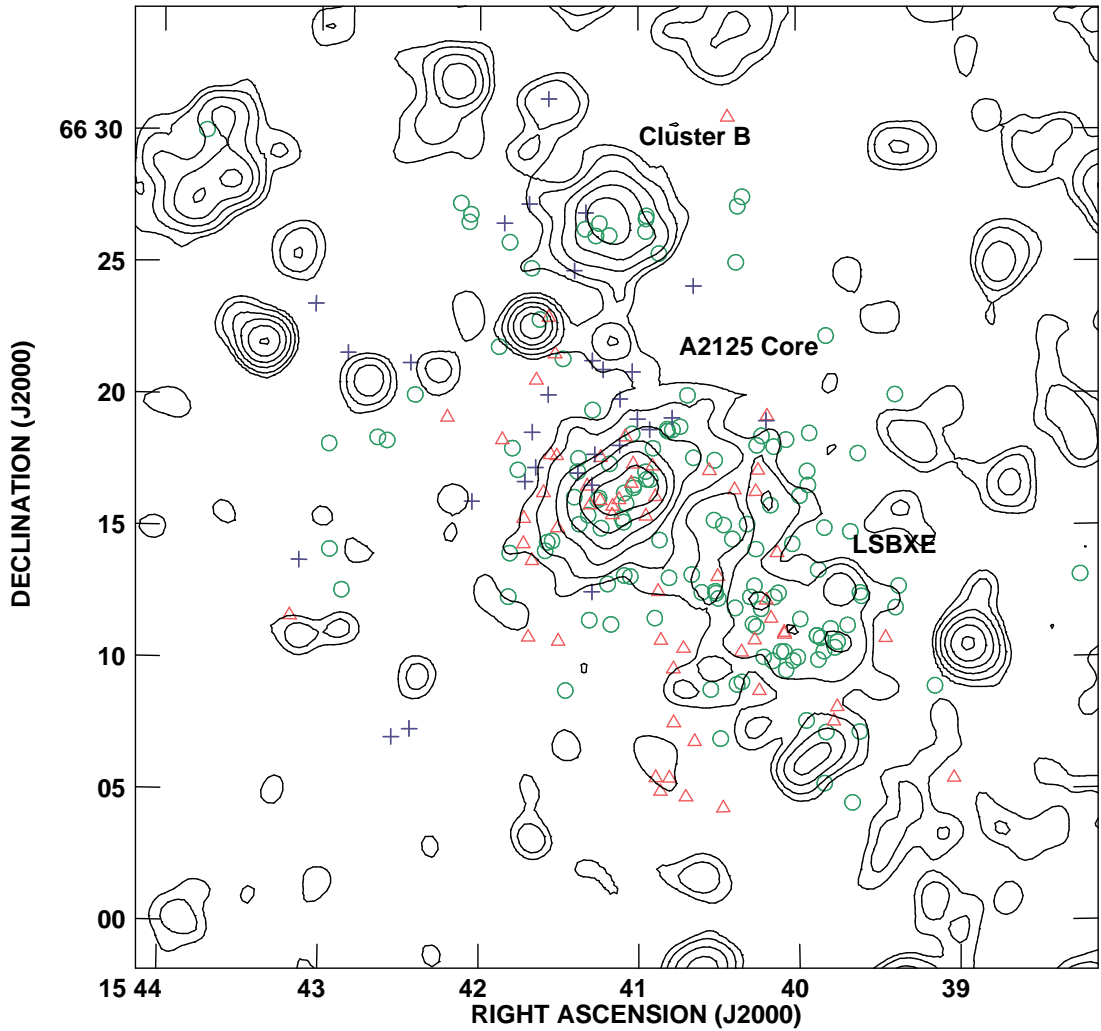}}
\caption{Distribution of Abell 2125 galaxies based on KMM fitting of three substructures, overlaid on X-ray contours from the smoothed {\it ROSAT} PSPC image. X-ray labels, as presented in \citet{wang1997}, are located above and to the right of the corresponding features. The red open triangles represent galaxies from the high velocity component, the blue pluses represent the low velocity component, and the green circles represent the main cluster (refer also to Figure \ref{fig-kmmvh} for the velocity histograms).\label{fig-dist}}
\end{figure}

\begin{figure}
\figurenum{5}
\epsscale{1.0}
\plotone{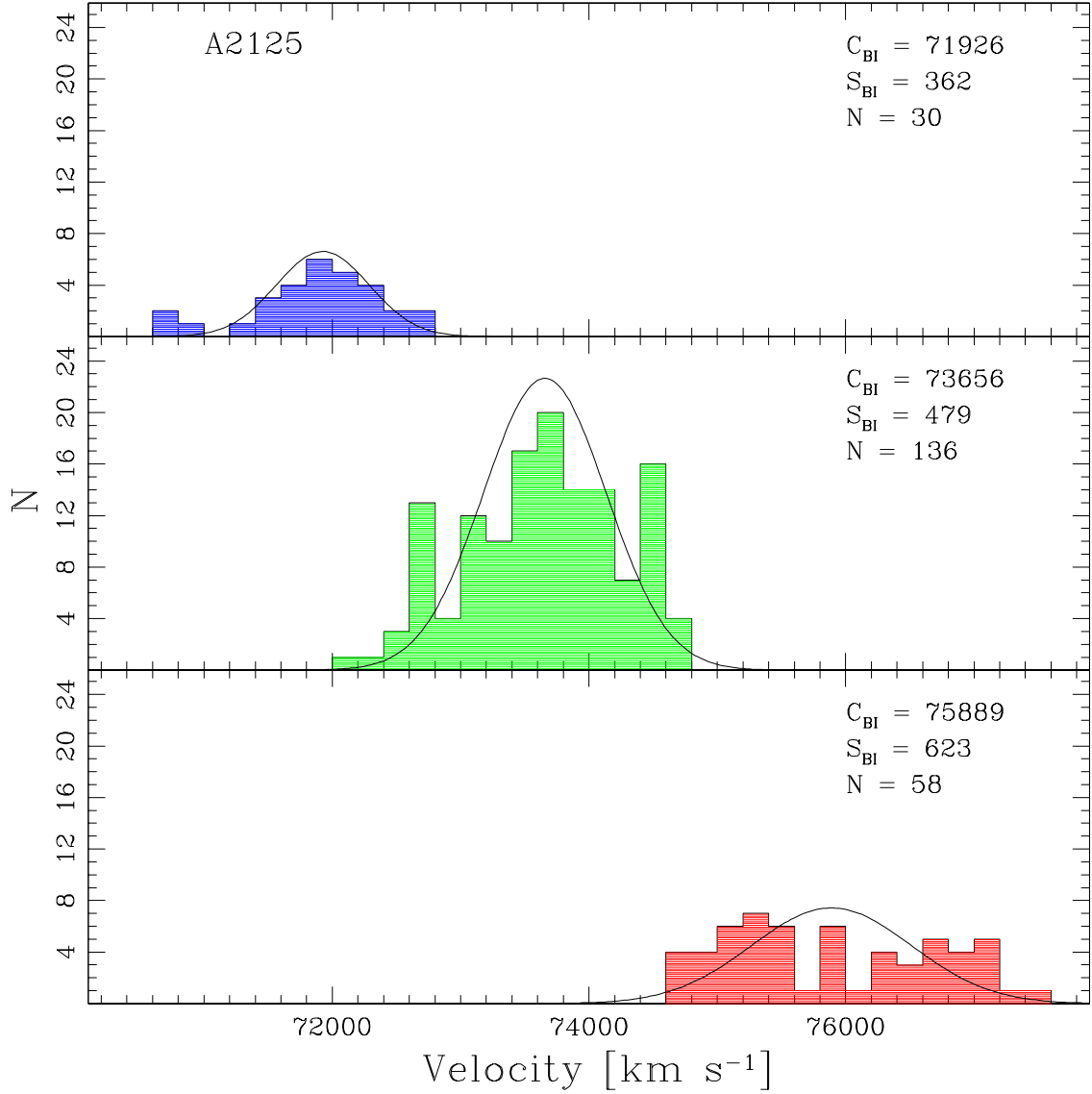}
\caption{Velocity histograms for the KMM fitting of three substructures (see Figure \ref{fig-dist} for positional information, using same color coding). Gaussians with mean and dispersion set to the biweight location and scale of each component are overplotted.\label{fig-kmmvh}}
\end{figure}

\end{onecolumn}

\end{document}